\documentclass[twocolumn]{aastex631}

\usepackage{amsmath}
\usepackage{needspace}
\usepackage{comment}

\newcommand{\CYTone}[1]{{#1}}
\newcommand{\CYTtwo}[1]{{#1}}
\newcommand{\CYTthree}[1]{{#1}}
\newcommand{\MEASUREMENT}[1]{{#1}}

\newcommand{\lvval}{2300_{-700}^{+1200}}
\newcommand{\mvval}{-3.56_{-0.37}^{+0.47}}
\newcommand{\rhalfval}{90_{-30}^{+30}}
\newcommand{\distval}{111_{-6}^{+9}}
\newcommand{\ellipval}{0.54_{-0.29}^{+0.19}}
\newcommand{\ageval}{12.37}
\newcommand{\vlosval}{170.03_{-1.41}^{+1.75}}
\newcommand{\vdispval}{2.85_{-1.31}^{+1.57}}
\newcommand{\dmmassval}{0.7_{-0.5}^{+1.2} \times 10^6 }
\newcommand{\mlratioval}{700_{-500}^{+1400}}
\newcommand{\fehval}{-2.39_{-0.13}^{+0.10}}
\newcommand{\fehdispval}{0.19_{-0.11}^{+0.14}}
\newcommand{\pmraval}{-0.06_{-0.20}^{+0.18}}
\newcommand{\pmdecval}{-0.22_{-0.17}^{+0.22}}
\newcommand{\apoval}{162_{-11}^{+109}}
\newcommand{\perival}{103_{-53}^{+11}}
\newcommand{\orbeccval}{0.58_{-0.03}^{+0.12}}
\newcommand{\orbEval}{-0.024 \ \pm 0.023}
\newcommand{\orbLzval}{ -4.16 \ \pm 4.71 }
\newcommand{\orbEpctval}{79}
\newcommand{\orbLzpctval}{84}
\newcommand{\rhoLeoval}{1.2_{-0.9}^{+3.8} \times 10^8}
\newcommand{\rhoMWval}{6.6_{-3.6}^{+38.8}  \times 10^5 }
\newcommand{\velgradval}{78^{+48}_{-49}}
\newcommand{\velgraduval}{40^{+25}_{-25}}
\newcommand{\velgradjackval}{47^{+56}_{-54}}
\newcommand{\velgradjackuval}{24^{+29}_{-28}}
\newcommand{\velgradbic}{1.3}

\newcommand{\veldispbic}{4.8}
\newcommand{\veldispbayes}{11.1}

\reportnum{\footnotesize FERMILAB-PUB-24-0358-LDRD-PPD}
\shorttitle{Discovery of Leo VI Ultra-Faint Dwarf Galaxy}
\shortauthors{DELVE Collaboration}

\begin{document}

\title{A Pride of Satellites in the Constellation Leo? Discovery of the Leo VI Milky Way Satellite Ultra-Faint Dwarf Galaxy with DELVE Early Data Release 3}

\correspondingauthor{Chin Yi Tan}
\email{chinyi@uchicago.edu}
\author[0000-0003-0478-0473]{C.~Y.~Tan}
\affiliation{Kavli Institute for Cosmological Physics, University of Chicago, Chicago, IL 60637, USA}
\affiliation{Department of Physics, University of Chicago, Chicago, IL 60637, USA}
\author[0000-0003-1697-7062]{W.~Cerny}
\affiliation{Department of Astronomy, Yale University, New Haven, CT 06520, USA}
\author[0000-0001-8251-933X]{A.~Drlica-Wagner}
\affiliation{Fermi National Accelerator Laboratory, P.O.\ Box 500, Batavia, IL 60510, USA}
\affiliation{Kavli Institute for Cosmological Physics, University of Chicago, Chicago, IL 60637, USA}
\affiliation{Department of Astronomy and Astrophysics, University of Chicago, Chicago, IL 60637, USA}
\author[0000-0002-6021-8760]
{A.~B.~Pace}
\affiliation{McWilliams Center for Cosmology \& Astrophysics, Carnegie Mellon University, 5000 Forbes Ave, Pittsburgh, PA 15213, USA}
\author[0000-0002-7007-9725]{M.~Geha}
\affiliation{Department of Astronomy, Yale University, New Haven, CT 06520, USA}
\author[0000-0002-4863-8842]{A.~P.~Ji}
\affiliation{Kavli Institute for Cosmological Physics, University of Chicago, Chicago, IL 60637, USA}
\affiliation{Department of Astronomy and Astrophysics, University of Chicago, Chicago, IL 60637, USA}
\author[0000-0002-9110-6163]{T.~S.~Li}
\affiliation{Department of Astronomy and Astrophysics, University of Toronto, 50 St. George Street, Toronto ON, M5S 3H4, Canada}
\author[0000-0002-6904-359X]{M.~Adam\'ow}
\affiliation{Center for Astrophysical Surveys, National Center for Supercomputing Applications, 1205 West Clark St., Urbana, IL 61801, USA}
\author[0000-0003-3312-909X]{D.~Anbajagane}
\affiliation{Department of Astronomy and Astrophysics, University of Chicago, Chicago, IL 60637, USA}
\affiliation{Kavli Institute for Cosmological Physics, University of Chicago, Chicago, IL 60637, USA}
\author[0000-0003-4383-2969]{C.~R.~Bom}
\affiliation{Centro Brasileiro de Pesquisas F\'isicas, Rua Dr. Xavier Sigaud 150, 22290-180 Rio de Janeiro, RJ, Brazil}
\author[0000-0002-3690-105X]{J.~A.~Carballo-Bello}
\affiliation{Instituto de Alta Investigaci\'on, Universidad de Tarapac\'a, Casilla 7D, Arica, Chile}
\author[0000-0002-3936-9628]{J.~L.~Carlin}
\affiliation{Rubin Observatory/AURA, 950 North Cherry Avenue, Tucson, AZ, 85719, USA}
\author[0000-0002-7887-0896]{C.~Chang}
\affiliation{Department of Astronomy and Astrophysics, University of Chicago, Chicago, IL 60637, USA}
\affiliation{Kavli Institute for Cosmological Physics, University of Chicago, Chicago, IL 60637, USA}
\author[0000-0001-5143-1255]{A.~Chaturvedi}
\affiliation{Department of Physics, University of Surrey, Guildford GU2 7XH, UK}
\author[0000-0002-7155-679X]{A.~Chiti}
\affiliation{Department of Astronomy and Astrophysics, University of Chicago, Chicago, IL 60637, USA}
\affiliation{Kavli Institute for Cosmological Physics, University of Chicago, Chicago, IL 60637, USA}
\author[0000-0003-1680-1884]{Y.~Choi}
\affiliation{NSF NOIRLab, 950 N. Cherry Ave., Tucson, AZ 85719, USA}
\author[0000-0002-1693-3265]{M.~L.~M.~Collins}
\affiliation{Department of Physics, University of Surrey, Guildford GU2 7XH, UK}
\author[0000-0001-9775-9029]{A.~Doliva-Dolinsky}
\affiliation{Department of Chemistry and Physics, University of Tampa, 401 West Kennedy Boulevard, Tampa, FL 33606, USA}
\affiliation{Department of Physics and Astronomy, Dartmouth College, Hanover, NH 03755, USA}
\author[0000-0001-6957-1627]{P.~S.~Ferguson}
\affiliation{Department of Physics, University of Wisconsin-Madison, Madison, WI 53706, USA}
\author[0000-0002-4588-6517]{R.~A.~Gruendl}
\affiliation{Department of Astronomy, University of Illinois, 1002 W. Green Street, Urbana, IL 61801, USA}
\affiliation{Center for Astrophysical Surveys, National Center for Supercomputing Applications, 1205 West Clark St., Urbana, IL 61801, USA}
\author[0000-0001-5160-4486]{D.~J.~James}
\affiliation{ASTRAVEO LLC, PO Box 1668, Gloucester, MA 01931}
\affiliation{Applied Materials Inc., 35 Dory Road, Gloucester, MA 01930}
\author[0000-0002-9269-8287]{G.~Limberg}
\affiliation{Kavli Institute for Cosmological Physics, University of Chicago, Chicago, IL 60637, USA}
\affiliation{Universidade de S\~ao Paulo, IAG, Departamento de Astronomia, SP 05508-090, S\~ao Paulo, Brazil}
\author[0000-0001-9438-5228]{M.~Navabi}
\affiliation{Department of Physics, University of Surrey, Guildford GU2 7XH, UK}
\author[0000-0003-3835-2231]{D.~Mart\'{i}nez-Delgado}
\affiliation{Instituto de Astrof\'{i}sica de Andaluc\'{i}a, CSIC, E-18080 Granada, Spain}
\author[0000-0002-9144-7726]{C.~E.~Mart\'inez-V\'azquez}
\affiliation{International Gemini Observatory/NSF NOIRLab, 670 N. A'ohoku Place, Hilo, Hawai'i, 96720, USA}
\author[0000-0003-0105-9576]{G.~E.~Medina}
\affiliation{Department of Astronomy and Astrophysics, University of Toronto, 50 St. George Street, Toronto ON, M5S 3H4, Canada}
\author[0000-0001-9649-4815]{B.~Mutlu-Pakdil}
\affiliation{Department of Physics and Astronomy, Dartmouth College, Hanover, NH 03755, USA}
\author[0000-0002-1793-3689]{D.~L.~Nidever}
\affiliation{Department of Physics, Montana State University, P.O. Box 173840, Bozeman, MT 59717-3840}
\affiliation{NSF NOIRLab, 950 N. Cherry Ave., Tucson, AZ 85719, USA}

\author[0000-0002-8282-469X]{N.~E.~D.~No\"el}
\affiliation{Department of Physics, University of Surrey, Guildford GU2 7XH, UK}
\author[0000-0001-5805-5766]{A.~H.~Riley}
\affiliation{Institute for Computational Cosmology, Department of Physics, Durham University, South Road, Durham DH1 3LE, UK}
\author[0000-0002-1594-1466]{J.~D.~Sakowska}
\affiliation{Department of Physics, University of Surrey, Guildford GU2 7XH, UK}
\author[0000-0003-4102-380X]{D.~J.~Sand}
\affiliation{Department of Astronomy/Steward Observatory, 933 North Cherry Avenue, Room N204, Tucson, AZ 85721-0065, USA}
\author[0009-0001-1133-5047]{J.~Sharp}
\affiliation{Department of Astronomy and Astrophysics, University of Chicago, Chicago, IL 60637, USA}
\author[0000-0003-1479-3059]{G.~S.~Stringfellow}
\affiliation{Center for Astrophysics and Space Astronomy, University of Colorado, 389 UCB, Boulder, CO 80309-0389, USA}
\author[0009-0007-0636-8876]{C.~Tolley}
\affiliation{Department of Astronomy, University of California, Berkeley, CA 94720-3411, USA}
\author[0000-0001-7211-5729]{D.~L.~Tucker}
\affiliation{Fermi National Accelerator Laboratory, P.O.\ Box 500, Batavia, IL 60510, USA}
\author[0000-0003-4341-6172]{A.~K.~Vivas}
\affiliation{Cerro Tololo Inter-American Observatory/NSF NOIRLab, Casilla 603, La Serena, Chile}

\collaboration{37}{(DELVE Collaboration)}




\begin{abstract}

We report the discovery and spectroscopic confirmation of an ultra-faint Milky Way (MW) satellite in the constellation of Leo. This system was discovered as a spatial overdensity of resolved stars observed with Dark Energy Camera (DECam) data from an early version of the third data release of the DECam Local Volume Exploration survey (DELVE EDR3). The low luminosity ($M_V = \mvval$\,; $L_V = \lvval L_\odot$), large size ($R_{1/2} = \rhalfval$\,pc), and large heliocentric  
distance ($D = \distval$\, kpc) are all consistent with the population of ultra-faint dwarf galaxies (UFDs). Using Keck/DEIMOS observations of the system, we were able to spectroscopically confirm nine member stars, while measuring a tentative \CYTthree{mass-to-light} ratio of  $\mlratioval M_\odot/L_\odot$ and a non-zero metallicity dispersion of $\sigma_{[\rm Fe/H]}=\fehdispval$, further confirming Leo VI's identity as an  UFD. While the system has a highly elliptical shape, $\epsilon = \ellipval$, we do not find any \CYTthree{conclusive} evidence that it is tidally disrupting. Moreover, despite the apparent on-sky proximity of Leo VI to members of the proposed Crater-Leo infall group,  its smaller heliocentric distance and inconsistent position in energy-angular momentum space make it unlikely that Leo VI is  part of the proposed infall group.  
\end{abstract}

\keywords{dwarf galaxies, surveys}

%% and \citet commands to identify citations. The citations are

\section{Introduction} \label{sec:intro}
 Ultra-faint dwarf galaxies (UFDs) are among the oldest, faintest ($M_V \gtrsim -7.7$; $L_V \lesssim 10^5 L_\odot$), most metal-poor ($\mbox{[Fe/H]} \lesssim -2$), and most dark-matter-dominated ($M_{\odot}/L_{\odot} > 100$) stellar systems known \citep{Simon:2019}. UFDs were first discovered in the Sloan Digital Sky Survey \citep[SDSS;][]{Willman:2005a, Willman:2005b} following the advent of CCD-based digital sky surveys. Subsequently, more recent surveys such as the Pan-STARRS-1 \citep[\CYTthree{PS1:}][]{Chambers:2016}, the Dark Energy Survey \citep[DES:][]{DES:2016}, and the DECam Local Volume Exploration Survey \citep[DELVE;][]{Drlica-Wagner:2021} have drastically expanded the known population of these faint resolved systems around the MW to more than 60 systems \citep[e.g.,][]{laevens:2015, Koposov15a, Bechtol:2015, Drlica-Wagner:2015, Drlica-Wagner:2020, Mau:2020, Cerny:2023c}.

The high dark matter content of UFDs makes them excellent laboratories for understanding the nature of dark matter. 
For example, the luminosity function of UFDs and their density profiles depend sensitively on the dark matter particle mass, thermal history, and self-interaction cross section \citep[e.g.,][]{Lovell:2014, Rocha:2013, Kaplinghat:2015, Bullock:2017}. Their internal dynamics are also sensitive to weak heating effects from the dark matter halo, allowing for potentially measurable effects on their stellar components  \citep{Brandt:2016, Penarrubia:2016}. UFDs are also excellent targets to search for Standard Model products coming from dark matter annihilation or decay due to their proximity, high dark matter content, and lack of high-energy astrophysical backgrounds \citep[e.g.,][]{Ackermann:2015zua, Bonnivard2015MNRAS.453..849B, Geringer-Sameth:2015, McDaniel:2023, Boddy:2024}.
Furthermore, the study of individual UFDs can yield insights into the processes of satellite accretion and tidal disruption around MW mass host galaxies. \CYTone{Due to their low luminosities, most of the UFD systems discovered thus far are resolved satellites of the MW and other nearby galaxies in the Local Volume (see \citealt{Simon:2019} and the reference therein.)}

%Although UFDs are thought to be the most common type of galaxies due to their low luminosities, most of the UFD systems discovered thus far are resolved satellites of the MW and other nearby galaxies in the Local Volume %\footnote{\CYTone{Examples of non Local Volume UFDs include  Virgo UFD-1 \citep{Jang:2014}, Fornax UFD-1 \citep{Lee:2017} and Ark227-UFD1 \citep{Conroy:2023}}}
%Sand et al. 2022, Mutlu-Pakdil et al. 2022, and Carlin et al. 2021.
 %Therefore, the discovery and characterization of UFDs surrounding the MW represents an observational frontier in near-field cosmology. 
%connecting the orbits of individual UFDs with dynamical indications of tidal disruption offers an opportunity to study the disruption of dark matter halos {\it in situ}.

In this paper, we present the discovery and confirmation of the UFD Leo VI, a low-luminosity, metal-poor MW satellite located at a heliocentric distance of $\sim$110 kpc. \CYTtwo{The paper also acts as an reference for the DELVE EDR3 dataset used to discover the system, which we describe in detail in Section~\ref{sec:delvedr3}. We use Section \ref{sec:delvesearch} to describe the  matched filter search methods used to discover Leo VI in DELVE EDR3 and  Section \ref{sec:fits} to present system's morphological properties obtained from follow-up DECam observations.} In Section \ref{sec:Keck}, we describe the line-of-sight velocity and metallicity measurements of the member stars of Leo VI obtained through follow-up Keck/DEIMOS data. We then discuss whether Leo VI could be tidally disrupting as well as \CYTone{its} potential associations with other Local Group systems in Section \ref{sec:discussion} \CYTone{and} summarize our results in Section \ref{sec:summary}.

\section{DELVE Early Data Release 3} \label{sec:delvedr3}

\CYTtwo{DELVE} is an ongoing observing program that uses DECam \citep{flaugher_2015_decam} on the 4-m Blanco Telescope at the Cerro Tololo Inter-American
Observatory (CTIO) in Chile to contiguously image the high Galactic latitude southern sky in the $g$, $r$, $i$, and $z$ bands \citep{Drlica-Wagner:2021, Drlica-Wagner:2022}. To date, DELVE has been allocated more than 150 nights to pursue three observational programs dedicated to studying ultra-faint satellite galaxies around the MW (DELVE--WIDE), the  Magellanic Clouds (DELVE--MC), and Magellanic analogs in the Local Volume (DELVE--DEEP). New DECam observations and public archival DECam data are self-consistently processed  using the DES Data Management pipeline \citep[DESDM:][]{Morganson:2018}. 

\CYTtwo{The forthcoming DELVE third data release combines} \CYTone{exposures from DELVE with other public DECam programs such as DES and the DECam Legacy Survey \citep[DECaLS:][]{Dey:2019}. Compared to the previous DELVE data release, the exposures were coadded to improve the depth and precision of photometric measurements. In this analysis, we use an object catalog from DELVE EDR3 which is an early internal version of the third DELVE data release containing only exposures in the Northern Galactic cap ($b > 10^\circ$).} \CYTtwo{While DELVE EDR3 is an internal data release, the full DELVE DR3 follows the same processing procedure and will be made publicly available in the future.} 

%The catalog contains multi-epoch source photometry obtained from the DECam data following the procedures developed for cosmology using the DES Y3 DEEP fields \citep[e.g.,][]{Hartley:2022, Everett:2022}. 
%To maintain consistency between the DESDM processing of DELVE \CYTone{E}DR3 DESDM processing pipeline and the pipeline used for DES analysis, we have minimized changes to our processing pipeline. 

The DELVE EDR3 data were self-consistently processed using the DESDM pipeline in the context of the DECam All Data Everywhere (DECADE) program at the \CYTone{National Center for Supercomputing Applications} (NCSA).  The configuration of the DESDM image detrending and coaddition pipeline closely follows the pipeline used to produce DES DR2 \citep{DES:2021}. We briefly summarize key aspects of the DESDM pipeline used to produce DELVE EDR3 here.

% This is summarizing Morganson 2018 section 3 & 4. Emphasize things that are different from the DECam Community Pipeline (Valdes et al. 2014)
% https://noirlab.edu/science/index.php/data-services/data-reduction-software/csdc-mso-pipelines/pl206
%\ADW{Boil down Section 3 and 4 of Morganson et al. 2018 into a few sentences here. Emphasize things that are not in the DECam CP: brighter fatter, sky PCA, satellite streak removal, etc.} 

Processing of individual DECam exposures was performed following the ``Final Cut'' pipeline described in \citet{Morganson:2018}. All exposures go through a preprocessing step, which includes crosstalk and overscan correction as well as bad pixel masking for each DECam CCD. In addition, we correct for the CCD nonlinearity with CCD-dependent lookup tables that convert the observed flux to the fitted model \citep{Bernstein:2017}. \CYTthree{Around bright sources, charges accumulating on strongly illuminated pixels will repel nearby charges, broadening the source's PSF in a phenomenon known as the ``brighter-fatter effect'' \citep[e.g.,][]{Antilogus:2014}.} To account for this effect, we use a CCD-dependent kernel derived from early DECam data by \citet{Gruen:2015}. We then apply flat field corrections using the raw, bias, dome flats, amplifier-specific conversion, and the non-linear correction to each CCD. For the bias and dome flat images, we used a set of ``supercals'' assembled for DES by combining bias and flats taken over several nights \citep[see Section 3.2 of][]{Morganson:2018}. The bias, dome flat images, and other image calibration data products were used corresponding to the nearest DES observing epoch.
%\CYT{Is that also true for non-DES images?}. \ADW{Yes}

At the time of observations, a world coordinate system (WCS) was added to the image  using the optical axis read from the telescope encoders and a fixed distortion map derived from the star flats. The WCS is then updated with an initial single-exposure astrometric solution calculated using \texttt{SCAMP} \citep{Bertin:2006}, with \textit{Gaia} DR2 \citep{Gaia:2018} used as the reference catalog without proper motion corrections.  The single-epoch astrometric accuracy is found to have a minimal value of $\approx$ 20 mas for DECam data taken close to the \textit{Gaia} 2015.5 epoch; however, it is found to increase before and after that date \citep{DES:2021}. 

To remove image artifacts, we mask the saturated pixels caused by bright stars, and the associated charge overflow in both the readout direction (``bleed trails'') and the CCD serial register (``edge bleeds''). We also perform cosmic ray masking using a modified version of the algorithm developed for the \CYTone{Legacy Survey of Space and Time}  \citep[LSST;][]{Juric:2017} and a satellite streak mask using an algorithm based on the Hough Transform \citep{Morganson:2018}.  

To account for the sky background light, we also fit and subtract the sky background from the whole DECam exposure using seasonally averaged PCA components and divide the image by the star flat \citep{Bernstein:2017}. The sky background model is produced from the raw bias, dome flat, and nightly star flat images and quantifies the differences between the dome flat and the response to astronomical flux. The point spread function (PSF) for each CCD image is obtained using \texttt{PSFEx} \citep{Bertin:2011}. Finally, the source catalog for each individual CCD image is then produced using \texttt{SourceExtractor} \citep{Bertin:1996} with a detection threshold of $\sim 3\sigma$ \citep{DES:2021}.

The DELVE EDR3 coadded images were assembled from a subset of DECam exposures that were publicly available as of 2022 December 5. The coadd input exposures were selected to reside in the northern Galactic cap ($b >  10\degr$) and to have an exposure time between 30 and 350 seconds.
Furthermore, we require that the images have a PSF full width at half maximum (FWHM) value of between 0'' $<$  FWHM $<$ 1.5'' \CYTtwo{and} that all exposures have an effective exposure time scale factor $t_{\rm eff} > 0.2$, where $t_{\rm eff}$ is calculated based on nominal values of the PSF FWHM, sky brightness, and transparency as described in \citet{Neilsen:2016}.  We also require that the input exposures have a good astrometric solution when matched to \textit{Gaia} DR2. This is achieved by requiring $>$ 250 astrometric matches, $\chi^2_{\rm astrom} < 180$, and an average difference of $\Delta_{\rm astrom} < 100$ mas.  As with \citet{Drlica-Wagner:2022}, to remove exposures with excess electronic noise or poor sky background estimation resulting in many spurious object detections, we also require that the number of objects detected in each exposure \CYTone{is less than the empirically determined limit of} $ 7.5 \times 10^5$. To improve the quality of the input exposures going into the coadding process, we further remove exposures that were identified as having suspect sky subtraction and/or astrometric fits based on selection criteria developed for DES \citep{DES:2021}.
%contains poor sky subtraction and suspect astrometric fit. 

\CYTtwo{Since }\CYTone{the public DECam exposures were taken for a wide variety of science purposes with different exposure times and filter distributions, there \CYTtwo{are} variations in survey depth and \CYTtwo{filter} coverage across the footprint. Therefore, to improve the uniformity of the dataset}, we select exposures to homogenize the cumulative effective exposure time across the DELVE footprint. 
This homogenization is performed by iteratively adding exposures to the coadd input list, starting with the exposure with the highest effective exposure time ($t_{\rm eff} \times t_{\rm exp}$). 
Exposures are iteratively added to the input list in order of effective exposure time unless $> 95\%$ of the exposure area is already covered by 15 or more exposures in the same band. \CYTtwo{The homogenization process removed 11\% of all the exposures and resulted in the standard deviation of the  effective exposure time ($t_{\rm eff} \times t_{\rm exp}$) across the survey area in $g$-band to drop from 1773s to 630s.} This selection results in the selection of \MEASUREMENT{61,425 exposures} \CYTtwo{in the $g,r,i,z$ bands} in the northern Galactic cap ($b > 10\degr$).  % I just check this is not done "if $>95\%$ of the exposure area has already been imaged by higher quality exposures with a summed effective exposure time $\sum t_{\rm eff} t_{\rm exp} > 270$\,s"

Individual CCD images are further checked for quality by automated algorithms and visual inspection \citep[e.g., looking for issues similar to those described in][]{Melchior:2016}. In particular, we identify images that are strongly affected by optical ghosting \citep{Kent:2013}, electronic noise variations, telescope motion (e.g., bad tracking, earthquakes, etc.), airplanes in the field of view, and other similar issues that lead to poor data quality. These quality checks are performed at both the exposure and individual CCD level, and the individual CCD images that contain artifacts are removed from the coadd input list. In total, 10,796 out of 6,440,109 CCD images were removed by this inspection.

Photometric calibration was performed by matching to the ATLAS RefCat2 reference catalog \citep{Tonry:2018}. ATLAS RefCat2 is an all-sky catalog that combines several surveys (i.e., \textit{Gaia}, PS1, SkyMapper, etc.). For the purposes of DELVE calibration, we utilize the PS1 measurements in the north (${\rm Dec.} > -30\degr$) and SkyMapper measurements in the south (${\rm Dec.} < -30\degr$).  The calibration followed the procedure described for DELVE DR2 \citep{Drlica-Wagner:2022}, but with an updated transformation procedure that finds the offsets required to convert the magnitudes for sources in different color bins and makes an interpolation for the intermediate color values 
\CYTthree{(Allam, S.\ S., Tucker, D.\ L., et al., in prep.).\footnote{\url{https://github.com/DouglasLeeTucker/TransformEqns}}} %\citep{Thanjavur:2021}
Separate interpolations were derived for the northern (PS1) and southern (SkyMapper) parts of the sky by matching the DES DR2 \texttt{WAVG\_MAG\_PSF} magnitudes to the corresponding ATLAS RefCat2  magnitudes to remove the $\sim$5 mmag offset detected in DELVE DR2 \citep{Drlica-Wagner:2022}. Photometric zeropoints for each DECam CCD were obtained by performing a $1\arcsec$ match between the DECam Final Cut catalogs and ATLAS RefCat2. The ATLAS RefCat2 measurements are transformed into the DECam system using interpolations, and the zeropoint is derived from the median offset required to match the ATLAS RefCat2 values. We repeated the calibration process for each CCD three times, with each iteration removing outlier sources where the magnitude difference between the zeropoint-calibrated sources and ATLAS RefCat2 values is $>3\sigma$ from the mean. The relative photometric uncertainty was validated through comparison to \textit{Gaia} EDR3 \CYTone{using transformation equations derived for DES DR2 \citep{DES:2021}} and by measurements of the width of the stellar locus following the procedures described in \citet{Drlica-Wagner:2022}. Calibrated single-epoch sources were found to agree with \textit{Gaia} magnitudes with a scatter of \CYTtwo{\MEASUREMENT{$\pm$4.2} mmag (estimated using half the width of the 68\% containment)}. %See delve/calibration/zps_replacement/v11/Paper%20values.ipynb

%\ADW{This is summarizing Morganson et al. 2018 Section 6}

The image coaddition process follows that described in Section 6 of \citet{Morganson:2018}. Coadded images are built in distinct rectangular tiles that have dimensions of $0.71 \times 0.71$ deg covered by 10,000 pixels with a pixel scale of 0.263 arcsec. Coadd images were constructed for tiles that had at least partial coverage in all four bands ($g$,$r$,$i$,$z$). To minimize the FWHM of the coadded PSF due to astrometric offsets between input exposures, we recompute a global astrometric solution for all CCD images provided as input to each coadd tile. For each tile, we use \texttt{SCAMP} \citep{Bertin:2006} to perform an astrometric refinement by simultaneously solving for the astrometry using the Final Cut catalogs for each input CCD and \textit{Gaia} DR2 as an external reference catalog. Using this process, the DES pipeline obtained residuals from the simultaneous astrometric fit with a standard deviation of $\sim$27 mas \citep{DES:2021}. We then use \texttt{SWARP} \citep{Bertin:2002} to resample the input images and produce the final coadd images for all four bands ($g,r,i,z$). In addition to the individual bands, we also produce a detection image, which is a coadd of the $r+i+z$ images using the \texttt{COMBINE$\_$TYPE=AVERAGE} procedure in \texttt{SWARP} \citep{DES:2021}.

Object detection is performed on the $r+i+z$ detection coadd image using \texttt{SourceExtractor} following the procedure described for DES DR2 \citep{DES:2021}.  Objects are detected when contiguous groups of 4 or more pixels exceed a threshold of 1.5$\sigma$, which has been found to correspond to a source detection threshold of $\sim 5\sigma$ \citep{DES:2021}. Initial photometric measurement is performed with \texttt{SourceExtractor} in ``dual image mode,'' using the detection image and the band of interest. 
Object astrometry comes from the windowed positions derived by \texttt{SourceExtractor} on the coadd images.  The DELVE EDR3 footprint contains 6,895 coadd tiles, which covers 7,737\,deg$^2$ with a median limiting magnitude of $g \sim 24.1$, $r\sim 23.6$, $i\sim 23.2$, $z \sim 22.5$ (estimated at ${\rm S/N} = 10$ in 2$\arcsec$ aperture from survey property maps derived from the single-epoch images that go into the coadds). \CYTone{For comparison, the median limiting magnitude of  DES DR2 assessed at the same S/N with the same technique is $g \sim 24.7$, $r\sim 24.4$, $i\sim 23.8$, $z \sim 23.1$ \citep{DES:2021}. 
In contrast, the DELVE DR2 limiting PSF magnitude at S/N =10 is $g \sim 23.5$, $r\sim 23.1$, $i\sim 22.7$, $z \sim 22.1$} \citep{Drlica-Wagner:2022}.

%\ADW{MEDS production would come maybe from Sevilla et al. 2021 (https://arxiv.org/abs/2011.03407) Section 3.3. Description of fitvd would come from Hartley et al. 2022.}

We perform multi-epoch photometric fitting following the procedures developed for cosmology analyses using the DES Year 3 DEEP fields \citep[e.g.,][]{Hartley:2022, Everett:2022}. We first create multi-epoch data structures \citep[MEDS;][]{Jarvis:2016, Zuntz:2018} consisting of cutouts centered on each object detected in the coadd images. The dimensions of the MEDS range in size from $32 \times 32$ to $256 \times 256$ pixels, and they are comprised of the individual constituent images that went into the coadd at the location of each detected object.
We perform multi-band, multi-epoch fitting using the \texttt{fitvd} tool \citep{Hartley:2022}, which is built on top of the core functionality of \texttt{ngmix} \citep{Sheldon:2014}.
We perform PSF model fits and bulge + disk model fits (BDF)  while masking neighboring sources (i.e., a ``single-object fit'', SOF, in the DES nomenclature). The PSF model fits are obtained by fitting the amplitude of the
individual-epoch PSF models, while the BDF fits consist of fitting a galaxy model with bulge and disk components that are S\'ersic profiles with fixed indices of $n = 4$ and $n = 1$, respectively. To reduce the degeneracies in the parameters, the relative effective radii of the bulge and disk components in the BDF fits are fixed to unity. %\CYT{talk about metacalibration?} 
The magnitudes referenced in this paper are from the \texttt{fitvd} PSF fit, which has been found to provide the best photometry for point-like sources \citep{DES:2021}.

%\ADW{Value Added columns}

We append several ``value added'' columns to the catalogs. Due to the increased depth of the DELVE EDR3 catalog relative to DELVE DR2 and the rapidly rising number density of faint background galaxies, the effects of star--galaxy misclassification can become more prominent in searches for resolved stellar systems. We perform star--galaxy separation using the sizes and signal-to-noise ratios of the sources measured by the multi-epoch \texttt{fitvd} fit \CYTtwo{following a classification procedure developed for DES Year 6 (Bechtol et al., in prep). The classifier assigns an integer object class (ranging from 0 being likely stars to 4 being likely galaxies)} to each source. Using DES data, a relatively pure sample of stars with $0\leq \texttt{EXT\_FITVD} \leq 1$ has been found to have a stellar efficiency \CYTone{(true positive rate)} of 90\%  with galaxy contamination  \CYTone{(false discovery rate)}  of 10\% when integrating over a magnitude range $19 \leq i \leq 23.5$, while a more complete stellar sample selected with $0\leq \texttt{EXT\_FITVD} \leq \CYTone{3}$ has a stellar efficiency of 96\% with galaxy contamination of 27\% over the same magnitude range. \CYTtwo{The relative stellar efficiency (and galaxy contamination) of the classification procedure starts to drop (increase) strongly with magnitude starting at {$i\sim$23}. }

%in the catalog based on their position in the pre-seeing bulge+disk model (\textit{BDF}) size and \textit{BDF} signal-to-noise ratio parameter space.

To calculate the extinction due to interstellar dust, we first obtain the value of $E(B-V)$ by performing a bi-linear interpolation to the \citet{Schlegel1998} maps at the location of each source in the catalog. 
The reddening correction for each band is then calculated using the fiducial interstellar extinction coefficients
from DES DR2  such that $A_b = R_b \times E(B-V)$ where $ R_g = 3.186$,
$R_r = 2.140$, $R_i = 1.569$, and $R_z = 1.196$ \citep{DES:2021}. 
As described in \citet{DES:2021}, these coefficients include a renormalization of the $E(B-V)$ extinction values ($N=0.78$) as suggested by \citet{Schlafly:2010} and \citet{Schlafly:2011} so that they can be used directly with the $E(B-V)$ values from the \citet{Schlegel1998} map.
In this paper, we denote extinction-corrected magnitudes with the subscript ``0''.

\section{The Discovery of Leo VI} \label{sec:delvesearch}

\begin{figure*}[ht]
\centering
\includegraphics[width=\linewidth]{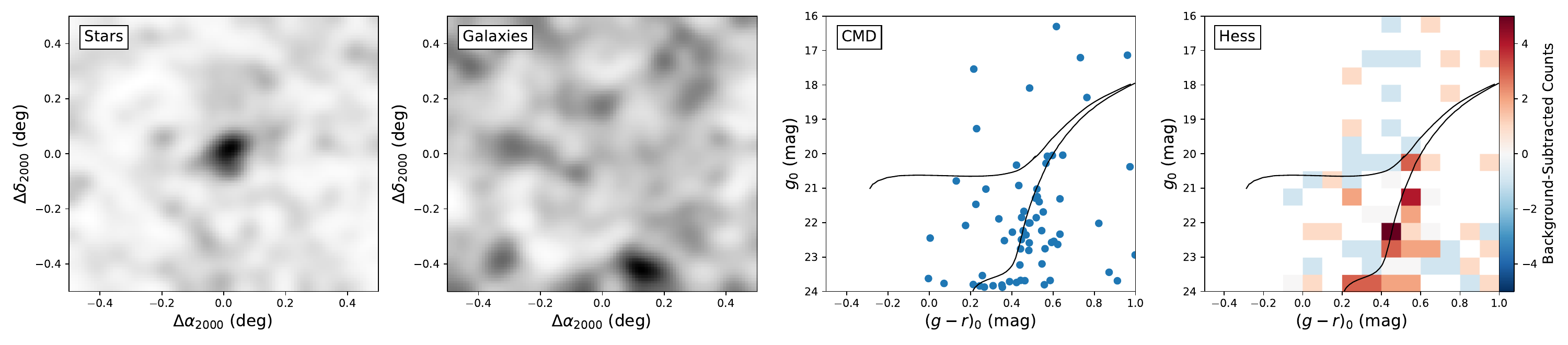}
%\vspace{-3cm}
\caption{\CYTone{Diagnostic plots for Leo VI constructed with the deeper follow-up DECam photometric data. {\it Left}:  Smoothed isochrone-filtered spatial density of stars in the region around the Leo VI. The system appears as a significant overdensity against a relatively constant background.  {\it Center Left}: Smoothed isochrone-filtered spatial density of galaxies. {\it Center Right}: Color–magnitude diagram (CMD) of stars that are within 4\arcmin \ of the centroid of the system.   {\it Right}: Color-magnitude Hess diagram of the foreground stars within  4\arcmin \ of the centroid of Leo VI minus the background stars in a concentric annulus with radius from 12\arcmin \ to 12.65\arcmin.  We use the
\texttt{ugali} best-fit old ($\tau \sim $13.3 Gyr), metal poor (Z = 0.0001, \mbox{[Fe/H]} $\sim -2.2$)  \texttt{PARSEC} isochrone with  distance modulus of $(m-M)_0 = 20.2$ for this figure.} \label{Figure:simple}}
\end{figure*}   

We search for resolved stellar systems in the DELVE EDR3 catalog using the \texttt{simple} matched-filter search algorithm \citep{Bechtol:2015,Drlica-Wagner:2020}.\footnote{https://github.com/DarkEnergySurvey/simple} This algorithm has been successfully used to discover more than twenty MW satellites to date \citep[e.g.][]{Bechtol:2015, Drlica-Wagner:2015, Mau:2020,Cerny:2021, Cerny:2023c}. To parallelize the search across the \CYTone{DELVE EDR3} footprint, we first \CYTone{partition the catalog into HEALPix} at the scale of NSIDE = 32 ($\approx$ 3.4 deg$^2$ per pixel). We then perform the search for each HEALPixelized catalog by combining the central catalog with catalogs from the 8 neighboring HEALPixels. 

In our \texttt{simple} satellite search, we select stars as sources with an object class of  $0\leq \texttt{EXT\_FITVD} \leq 1$. To reduce the effect of foreground contaminant stars, we only select stars that are consistent with an old
($\tau$ = 12 Gyr), metal poor ($Z$ = 0.0001, \mbox{[Fe/H]} $\sim -2.2$) \texttt{PARSEC} isochrone \citep{Bressan:2012, Chen:2014, Tang:2014, Chen:2015}. This is done by selecting stars that satisfy the following conditions  $\Delta(g-r)_0 < \sqrt{0.10^2 + \sigma^2_g + \sigma^2_r}$, \CYTone{ where $\sigma_g$ and  $\sigma_r$ are the uncertainties of the $g$ and $r$ magnitudes of the individual stars}. We perform the search multiple times as we scan the distance modulus of the isochrone in a range of $16.0 \leq m - M \leq 24.0$ mag at intervals of 0.5 mag. After the selection cuts, we smoothed the filtered stellar density field with a  $1\arcmin$  Gaussian kernel. We then identify overdensities in the stellar density field by iteratively increasing the density threshold until only ten peaks remain. We computed the Poisson significance of the overdensities relative to the background stellar density (calculated using stars within a distance between 0.3 deg and 0.5 deg from the overdensity). We repeat the same procedure \CYTtwo{with the $g, i$ band pair} and select candidates with detection significance above a significance threshold of $5.5\sigma$ in both
\CYTtwo{$g, r $ bands and $g, i$ bands} \CYTtwo{(similar to in \citealt{Cerny:2023c})}. For each dwarf galaxy candidate, we produce a diagnostic plot containing the smoothed stellar density and color-magnitude diagram of the candidate, \CYTone{similar to Figure \ref{Figure:simple}.}

%\subsection{Discovery of Leo VI} \label{sec:discovery}
\CYTone{From the diagnostic plots, we identified a promising} candidate stellar system near $(\alpha_{J2000},\delta_{J2000}) = (171.05^{\circ},24.87^{\circ})$ at a \texttt{simple} detection significance of  \MEASUREMENT{$7.0 \sigma$} in the \CYTtwo{$g, r $ bands} and \MEASUREMENT{$7.1 \sigma$} in the \CYTtwo{$g, i $ bands}. \CYTone{Our analysis using follow-up observations of the system suggest it to be a newly discovered MW satellite UFD (see Section \ref{sec:summary})}. Therefore, following historical convention, we refer to this system as Leo VI.\footnote{After Leo I, Leo II \citep{Harrington50}, Leo A \citep[Leo III,][]{Zwicky42}, Leo IV \citep{Belokurov07} and Leo V \citep{Belokurov08}.}

Due to the lack of spatial coverage around Leo VI and the relatively shallow depth of the DELVE EDR3 data in this region,  we obtained additional follow-up DECam imaging of the candidate so that more accurate morphological fits could be obtained. The follow-up observations consist of 3 $\times$ 300-second imaging in \textit{g} and \textit{i} band taken on \CYTone{2023 June 17}. To increase the depth of the follow-up imaging, we coadd the new exposures with archival DECam exposures around the candidate using the same pipeline as the DELVE EDR3 catalog. However, we have reduced the $t_{\rm{eff}}$ cut from $t_{\rm eff}>0.2$ to $t_{\rm eff}>0.15$ to include the new follow-up exposures that were taken during less-than-ideal observing conditions. Nevertheless, \CYTone{the looser  $t_{\rm{eff}}$ cut still excludes one of the } 300-second \textit{i} band follow-up exposures due to its low effective exposure time scale factor, $t_{\rm eff}\approx0.07$, caused by cloudy and bright observing conditions. 

Compared to the initial DELVE EDR3 data, the new imaging has an increased the depth by 0.3 mag in the \textit{g}-band and 0.4 mag in the \textit{i}-band, leading \CYTone{the \texttt{simple} detection significance to} increase to \MEASUREMENT{$9.8 \sigma$} in the \CYTtwo{$g, r $ bands} and \MEASUREMENT{$9.7 \sigma$} in the \CYTtwo{$g, i $ bands}.

\begin{figure*}[ht]
\centering
\includegraphics[width=0.85\linewidth]{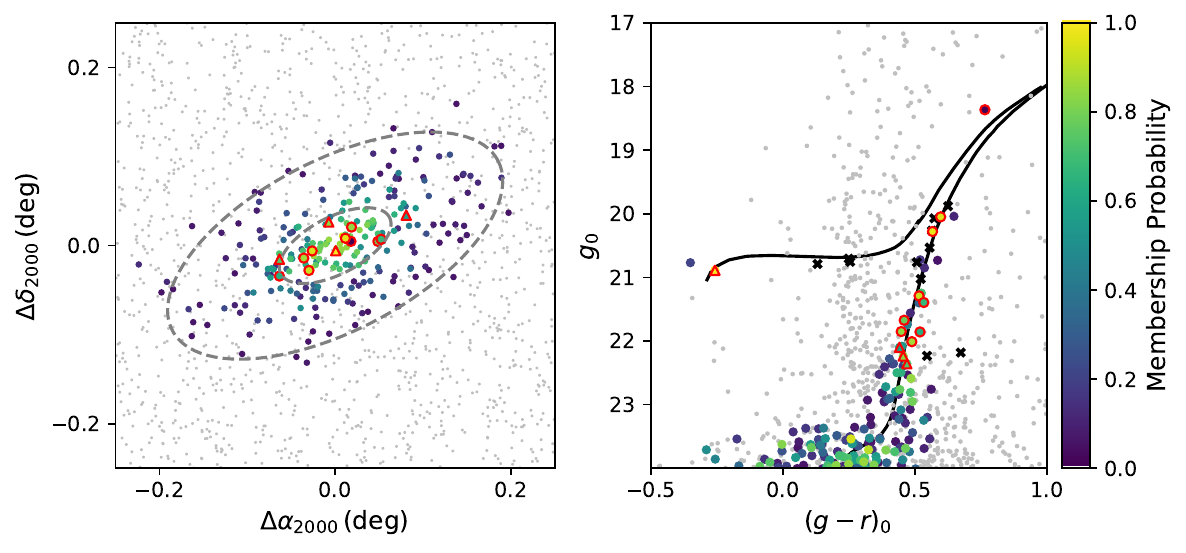}
%\includegraphics[width=\linewidth]{plots/Fig1.pdf}

%\vspace{-3cm}
\caption{
{\it Left}:   Map of likely member stars in the vicinity of the Leo VI system, with stars colored by their \texttt{ugali} membership probability. Stars with a lower \texttt{ugali} membership probability ($p < 0.05$) are shown in grey. 
{\it Right}:  Color–magnitude diagram of likely member stars in the system, again colored by their \texttt{ugali} membership probability. In both plots, \CYTthree{ stars with a red outline represent spectroscopically confirmed member stars  (circles) and candidate member stars (triangles) of Leo VI (Section \ref{sec:Keck}).} \CYTone{ The black crosses denotes sources that have confirmed to be non-members of Leo VI either through their inconsistent spectroscopic radial velocity, high \textit{Gaia} proper motion or \textit{Gaia} classification as likely quasar-stellar objects. The black curve represents the \texttt{ugali} best-fit \texttt{PARSEC} isochrone ($\tau \sim $13.3 Gyr, $Z$ = 0.0001, $(m-M)_0 = 20.2$).}}
\label{Figure:ugali}
\end{figure*}

\begin{deluxetable*}{lccc}
\tabletypesize{\scriptsize}
\tablewidth{0pt} 
%\tablenum{1}
\tablecaption{Measured and derived parameters of Leo VI. Details of each parameter can be found in their corresponding sections. \label{table:Leo6_param} }
\tablehead{
\colhead{Parameter} & \colhead{Description}& \colhead{Value} & \colhead{Units}   } 
\startdata 
%%%%%%%%%%%%%%%%%%%%%%%%%%%%%%%%%%%
\multicolumn{2}{l}{\textbf{Morphological Fits (Section 4)}}&& \\
$\alpha_{J2000}$& Right Ascension of Centroid & $171.077_{-0.013}^{+0.016}$ & deg \\ 
$\delta_{J2000}$& Declination of Centroid & $24.874_{-0.011}^{+0.008}$ & deg \\ 
$a_h$ &  Angular Semi-Major Axis Length & $4.17_{-1.33}^{+1.14}$ & arcmin  \\
$a_{1/2}$&  Physical Semi-Major Axis Length & $140_{-40}^{+40}$ &  pc  \\ 
$R_h$&  Azimuthally-Averaged Angular Half-Light Radius & $2.84_{-0.9}^{+0.78}$ &  arcmin \\
$R_{1/2}$&  Azimuthally-Averaged Physical Half-Light Radius & $ 90_{-30}^{+30}$ & pc  \\ 
$\epsilon$&  Ellipticity & $0.54_{-0.29}^{+0.19}$ &  -   \\
P.A.&  Position Angle of Major Axis (East of North) & $63_{-20}^{+13}$ & deg \\ 
$(m-M)_0$&  Distance Modulus$^a$& $20.23_{-0.12}^{+0.17}$ & mag\\
$D_\odot$&  Heliocentric Distance$^a$ & $111_{-6}^{+9}$ & kpc  \\ 
$\tau$&  Age$^b$ & $>$ 12.37  & Gyrs  \\
$M_V$&  Absolute (Integrated) $V$-band Magnitude & $-3.56_{-0.37}^{+0.47}$ &  mag   \\ 
$L_V$&  $V$-band  Luminosity & $2300_{-700}^{+1200}$ &  $L_\odot$  \\
$M_*$& \CYTthree{Stellar Mass (Assuming $M_*/L_V=2$)} & $4600_{-1400}^{+2300}$ & $M_\odot$  \\
\hline
\multicolumn{2}{l}{\textbf{Stellar Kinematics and Metallicities (Section 5)}}&& \\
$N_{\rm spec}$& Number of Spectroscopically Confirmed Members  Stars &  9   &  - \\
$v_{\rm hel}$& Heliocentric Radial Velocity & $170.03_{-1.41}^{+1.75}$ &  km s$^{-1}$  \\ 
$\sigma_{v}$& Line-of-Sight Velocity Dispersion & $2.85_{-1.31}^{+1.57}$ &  km s$^{-1}$  \\ 
$M_{1/2}$& Dynamical Mass within half-light radius$^c$ & $0.7_{-0.5}^{+1.2} \times 10^6 $ &  $M_\odot$  \\ 
$M_{1/2}/L_{V,1/2}$& Mass-to-Light Ratio within half-light radius$^c$ & $700_{-500}^{+1400}$ &  $M_\odot/L_\odot$  \\
$[\rm Fe/H]_{\rm spec}$& Mean Spectroscopic Metallicity & $-2.39_{-0.13}^{+0.10}$ &  dex \\ 
$\sigma_{[\rm Fe/H]}$& Metallicity Dispersion &  $0.19_{-0.11}^{+0.14}$ &  dex \\ 
\hline
\multicolumn{2}{l}{\textbf{Proper Motion, Orbits \& J-Factor (Section 6)}}&& \\
$\mu_{\alpha * }$& Proper Motion Right Ascension & $-0.06_{-0.20}^{+0.18}$ & mas yr$^{-1}$ \\
$\mu_{\delta}$& Proper Motion Declination & $-0.22_{-0.17}^{+0.22}$ & mas yr$^{-1}$ \\
$d_{GC}$& Galactocentric Distance & $114_{-7}^{+7}$ & kpc \\
$r_{\rm apo}$& Orbital Apocenter &  $162_{-11}^{+109}$ & kpc \\
$r_{\rm peri}$& Orbital Pericenter& $103_{-53}^{+11}$ & kpc \\
$e$& Orbital Eccentricity & $0.58_{-0.03}^{+0.12}$ & - \\
$\log_{10}J(0.5^{\circ})$& J-factor within a solid angle of 0.5$^{\circ}$ & $17.3_{-1.0}^{+0.9}$  &  GeV$^2$cm$^{-5}$ \\
%%%%%%%%%%%%%%%%%%%%%%%%%%%%%%%%%%%%%%%%%%%%%%%
\enddata
\tablenotetext{a}{Following \citet{Drlica-Wagner:2015}, we added in quadrature a 0.1 mag systematic uncertainty  to the distance modulus measurement to account for the uncertainties in the isochrone modeling.  }
\tablenotetext{b}{The posterior distribution peaked near $\tau$ = 13.5 Gyrs corresponding to our oldest available PARSEC isochrone, we therefore quote the  68\% Bayesian lower limit }
\tablenotetext{c}{\CYTtwo{ Based on single epoch velocities and}
assuming dynamical equilibrium.}
\end{deluxetable*}
\section{Morphological Fits} \label{sec:fits}
To obtain an estimate of the morphological properties of Leo VI and its stellar population, we use the maximum-likelihood-based Ultra--faint GAlaxy LIkelihood toolkit (\texttt{ugali}, \citealt{Bechtol:2015, Drlica-Wagner:2020}) \CYTone{on the followup DECam observations of the system}. We model the stellar density profile of the system with an elliptical Plummer profile \citep{Plummer:1911}, with the free parameters described by the centroid coordinates ($\alpha_{J2000},\delta_{J2000}$), angular semi-major axis length, $a_h$, ellipticity, $\epsilon$, and the position angle (P.A.) of the major axis (defined East of North). We then model the magnitudes and colors of the possible member stars with  \texttt{PARSEC} isochrone model \citep{Bressan:2012, Chen:2014, Tang:2014, Chen:2015}  with free parameters being the distance modulus, $(m-M)_0$, age, $\tau$, and the metallicity, $Z_{\rm phot}$, of the system. We also fit another free parameter: stellar richness, $\lambda$, which normalizes the
 \CYTtwo{total number of stars in the system} \citep{Bechtol:2015, Drlica-Wagner:2020}. % \CYT{Talk about fixing probability of member stars confirmed from spectra}  

We use the Markov Chain Monte Carlo sampler \texttt{emcee} \citep{Foreman_Mackey:2013} to simultaneously fit all the stellar density profiles and isochrone parameters in addition to the stellar richness \CYTthree{(posterior distributions are shown in Figure \ref{Figure:AppendixA})}. For star-galaxy separation, we loosen the stellar classification in our \texttt{ugali} analysis by defining stars to be sources with an object class of  $0\leq \texttt{EXT\_FITVD} \leq 3$ to include as many possible Leo VI member \CYTone{stars} as possible in our analysis. 

Table \ref{table:Leo6_param} shows the values and uncertainties of the stellar density profile and isochrone parameters obtained from \texttt{ugali}. The estimates of parameters are obtained from the median of the marginalized posteriors, while the 1$\sigma$ uncertainties are obtained using the 16th and 84th percentiles. \CYTone{ We find that posterior distribution for both the age of the system and the metallicity of the system peaked near the $\tau = 13.5$ Gyrs and $Z_{\rm phot}$ = 0.0001, \mbox{[Fe/H]} $\sim -2.2$ which represents oldest and most metal-poor isochrone in the \cite{Bressan:2012} library. \CYTtwo{We discuss spectroscopic measurements of the metallicity in Section \ref{sec:metallicity}.}}

Table \ref{table:Leo6_param} also shows Leo VI properties derived from the fitted parameters. For example, we can obtain the azimuthally-averaged angular half-light radius (defined as $R_h$ = $a_h \sqrt{1-\epsilon}$), the equivalent physical semi-major axis length (in parsec), $a_{1/2}$, and azimuthally-averaged physical half-light radius (in parsec), $R_{1/2}$. Using the prescription defined \cite{Martin2008}, we obtain the absolute $V$-band magnitude, $M_V$, from the isochrone and use it to derive the $V$-band luminosities ($L_V$). We also derive the stellar mass ($M_*$) of the system by \CYTthree{assuming a stellar-mass-to-light ratio of  $M_*/L_V=2$.} 

To investigate the robustness of the fits, we rerun \texttt{ugali} with
with different magnitude limit masks ranging from $g < 22.5$ to $g < 24.5$ mag at intervals 0.5 mag. We find that the fit results for all fitted parameters are consistent within 1$\sigma$ with the uncertainty being generally smaller when using the deeper mask. 

 \CYTtwo{For each star, the \texttt{ugali} pipeline assigns a probability that the star is a member of the dwarf galaxy based on its spatial position, photometric properties, and local imaging depth assuming a given model that includes a putative dwarf galaxy and the local stellar field population \citep{Bechtol:2015, Drlica-Wagner:2020}.}
 \CYTone{We plot the spatial distribution of stars in a small region around Leo VI, with stars colored by their \texttt{ugali} membership probability in the left plot of Figure \ref{Figure:ugali}. While in the right plot of same figure,  we show a color-magnitude diagram of the system with its stars colored by their \texttt{ugali} membership probability and also show the best-fit  \texttt{PARSEC} isochrone model \citep{Bressan:2012, Chen:2014, Tang:2014, Chen:2015}.}
 
Being excellent standard candles \citep{Catelan:2015}, the presence of  RR Lyrae (RRL) stars \CYTone{in Leo VI could be used to obtain an independent and more accurate  distance estimate of the system} \citep{Martinez-Vazquez:2019}. We search the \textit{Gaia} DR3 \citep{Clementini:2023} and  PS1 RRL \citep{Sesar:2018} catalog for RRL stars around Leo VI, but find no RRL stars within 0.2 degrees of the centroid of Leo VI. \CYTone{This is expected as the location of the Horizontal Branch at $g \sim$ 21 is around the magnitude limit of \textit{Gaia} and thus  deeper multi-epoch} observations are needed to detect potential RRL stars.

\begin{deluxetable*}{cccccccccc}
\label{table:keck_members}
\tablecaption{Properties of spectroscopically confirmed member stars and candidate member stars of Leo VI ordered by decreasing Keck/DEIMOS spectrum S/N. \CYTthree{ The top section lists confirmed member stars with line-of-sight velocity errors of $\epsilon_{v_{\rm hel}} < 10 {\rm{ \ km \ s}}^{-1}$, while the bottom section contains candidate member stars with less precise velocity data ($\epsilon_{v_{\rm hel}}>10 {\rm{ \ km \ s}}^{-1}$).} Details of each parameter can be found in Section \ref{sec:Keck}.}
\tablehead{Star Name & RA & DEC & $g_0$ & $r_0$ & S/N & $v_{\rm hel}$ &  $\Sigma$ EW CaT & [Fe/H]  & Type \\
     & (deg) & (deg) & (mag)&  (mag) & & (km s$^{-1}$)  & (\r{A})  & (dex) & }
\startdata
%%%%%%%%%%%%%%%%%%%
\textit{Gaia} DR3 3993822846542975360 & 171.096 & 24.878 & 18.4 & 17.6 & 66.6 & 168.9 $\pm$ 1.1 & 3.53 $\pm$ 0.21 & -2.60 $\pm$ 0.10 & RGB \\ 
\textit{Gaia} DR3 3993822949622182272 & 171.048 & 24.868 & 20.1 & 19.4 & 22.0 & 164.6 $\pm$ 1.6 & 2.85 $\pm$ 0.28 & -2.50 $\pm$ 0.13 & RGB \\ 
\textit{Gaia} DR3 3993823052701408128 & 171.090 & 24.882 & 20.3 & 19.7 & 18.9 & 174.0 $\pm$ 1.7 & 2.34 $\pm$ 0.37 & -2.69 $\pm$ 0.16 & RGB \\ 
\textit{Gaia} DR3 3993822502945568128 & 171.044 & 24.846 & 21.3 & 20.8 & 7.7 & 170.3 $\pm$ 4.0 & 3.03 $\pm$ 0.37 & -2.14 $\pm$ 0.17 & RGB \\ 
Leo VI J112432.23+245250.95 & 171.134 & 24.881 & 21.4 & 20.9 & 7.7 & 172.3 $\pm$ 3.4 & 3.51 $\pm$ 0.44 & -1.89 $\pm$ 0.19 & RGB \\ 
Leo VI J112431.27+245242.21 & 171.130 & 24.878 & 21.7 & 21.2 & 5.8 & 175.5 $\pm$ 4.5 & 2.46 $\pm$ 0.53 & -2.33 $\pm$ 0.23 & RGB \\ 
Leo VI J112423.32+245340.09 & 171.097 & 24.894 & 21.9 & 21.4 & 4.8 & 178.6 $\pm$ 6.6 & 3.26 $\pm$ 0.77 & -1.90 $\pm$ 0.34 & RGB \\ 
Leo VI J112401.61+245021.90 & 171.007 & 24.839 & 21.9 & 21.3 & 3.9 & 168.4 $\pm$ 8.3 & 3.16 $\pm$ 0.59 & -1.95 $\pm$ 0.26 & RGB \\ 
Leo VI J112408.96+245134.74 & 171.037 & 24.860 & 22.0 & 21.5 & 3.7 & 167.0 $\pm$ 7.0 & 2.05 $\pm$ 0.65 & -2.48 $\pm$ 0.30 & RGB \\ 
\hline
Leo VI J112418.58+245204.63 & 171.077 & 24.868 & 20.9 & 21.1 & 4.3 & 179.2 $\pm$ 15.1 & - & - & BHB \\ 
Leo VI J112439.81+245428.12 & 171.166 & 24.908 & 22.1 & 21.7 & 3.5 & 183.6 $\pm$ 10.3 & - & - & RGB \\ 
Leo VI J112416.48+245400.62 & 171.069 & 24.900 & 22.2 & 21.8 & 3.4 & 172.0 $\pm$ 19.0 & - & - & RGB \\ 
Leo VI J112401.62+245130.33 & 171.007 & 24.858 & 22.4 & 21.9 & 2.5 & 155.0 $\pm$ 12.7 & - & - & RGB \\ 
%%%%%%%%%%%%%%%%%%%%%%%%%
\enddata
%\tablecomments{Comments go here.}
\end{deluxetable*}

\section{Stellar Kinematics and Metallicities} \label{sec:Keck}
%\subsection{Keck Observations}
\subsection{Keck Observations} 
To confirm that Leo VI is a physically bound stellar system and not a chance arrangement of MW stars, we took follow-up spectroscopic observations of the potential member stars with the DEep Imaging Multi-Object Spectrograph (DEIMOS) mounted on the Keck II telescope \citep{Faber:2003}.

 We obtained 1.0 hour of DEIMOS observations of Leo VI through on the night of \CYTone{2024 February 14}.\footnote{Due to an earthquake-induced motor failure, the Keck II dome couldn't rotate during our run, limiting our total exposure time to when Leo VI transited the fixed azimuth window set by the dome slit. All of our exposures suffered from vignetting from the dome, resulting in lower $S/N$ but no other consequences of note.}
These observations used a single multi-object mask with slits of width 0.7'' and minimum length 4.5''. Targets were selected primarily based on the photometric probabilities provided by a \texttt{ugali} fit to the follow-up DECam imaging, as well as the astrometric information provided by \textit{Gaia}. 
 We used the DEIMOS 1200G grating and OG550 order blocking filter; this configuration provides $R \approx 6000$ across a wavelength range spanning H$\alpha$, the telluric A-band, and the Calcium Triplet region ($\sim 6500-9000 \rm \AA$). All exposures were reduced using the official Keck Data Reduction Pipeline found in the \texttt{PypeIt} software package \citep{Prochaska:2020}, with \texttt{PypeIt}'s default flexure correction disabled. Wavelength calibration was performed using XeNeArKr arcs, and flat-fielding used internal quartz flats.

\subsection{Line-of-Sight Velocities and Stellar Membership} \label{sec:veldisp}

\begin{figure*}[ht]
\centering
\includegraphics[width=\linewidth]{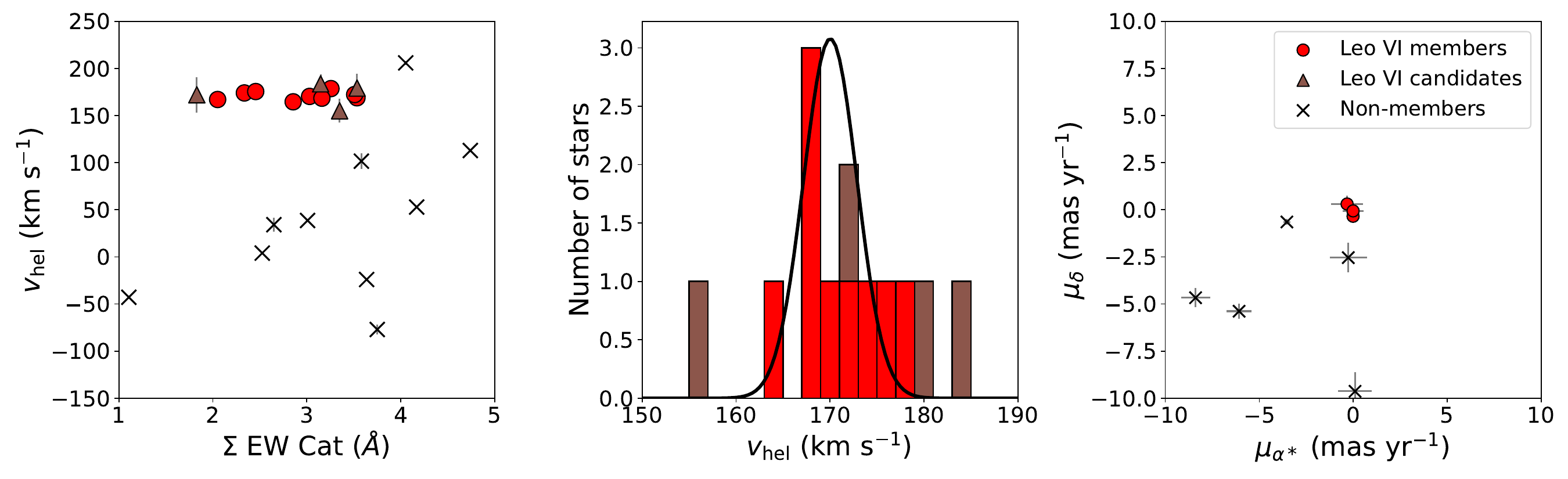}
%\vspace{-3cm}
\caption{
{\it Left}: The  Equivalent Widths (EWs) of the CaT lines vs.\ the measured line-of-sight (radial) velocity of the target stars. The plot shows a prominent excess of stars around 155 kms$^{-1}$ $\leq v_{\rm hel} \leq $ 185 kms$^{-1}$, which we assumed to be \CYTthree{member stars (red circles) and candidate member stars (brown triangles)} of Leo VI. The grey bars show the uncertainties of the line-of-sight (radial) velocity measurement.
{\it Middle}: Histogram of the measured heliocentric radial velocity of the Leo VI {member stars and candidate member} stars overlaid with the best-fit velocity dispersion model (See Section \ref{sec:veldisp}). \CYTthree{We note that the histograms does not capture the large uncertainties in the velocity measurements for the low S/N  stars.}
{\it Right}: Proper motions of \CYTthree{spectroscopic targets} that are bright enough ($g\lesssim 21$) to be measured in \textit{Gaia} DR3. Three of the spectroscopically confirmed member stars that also contain \textit{Gaia} proper motion measurements (red circles)  are found to be clustered near $\mu_\delta, \mu_\alpha \cos(\delta) \sim (0,0)$\, mas/yr.}
\label{Figure:Leo6Keck}
\end{figure*}

We obtain the line-of-sight velocities of the potential member stars using the \texttt{DMOST} package  (M.~Geha et al., in prep),\footnote{\url{https://github.com/marlageha/dmost}} which is a dedicated measurement pipeline for spectra obtained from DEIMOS 1200G grating.
The \texttt{DMOST} package measures line-of-sight velocities by forward modeling the spectrum of a star using both a \texttt{PHOENIX} stellar atmosphere library template \citep{Husser:2013} and a \texttt{TelFit} telluric absorption spectrum template \citep{Gullikson:2014}; the latter is used to correct for \CYTtwo{wavelength} shifts induced by the miscentering of stars within their slits. \CYTthree{The final velocity uncertainty is calculated by adding in quadrature the scaled statistical uncertainty and a $1.1 \ {\rm{ km \ s}}^{-1}$ uncertainty floor}. Further details about the velocity measurement procedure implemented within \texttt{DMOST} can be found in M.~Geha et al. (in prep.).

Starting from our initial target sample of \MEASUREMENT{44} stars, we were able use the pipeline to obtain line-of-sight velocity measurements for \MEASUREMENT{25} stars. We can see from the left subplot of Figure \ref{Figure:Leo6Keck} an excess of \MEASUREMENT{13} stars with a line-of-sight velocity, $v_{\rm hel}$, in the range of 155 kms$^{-1}$ $\leq v_{\rm hel} \leq $ 185 kms$^{-1}$, which we infer to be \CYTthree{possible} member stars of Leo VI, thus confirming the nature of Leo VI as a gravitationally-bound system. \CYTthree{We find \MEASUREMENT{9} spectroscopically confirmed member stars with line-of-sight velocity measurements that are precise enough to be included in the systemic analysis ($\epsilon_{v_{\rm hel}} < 10 {\rm{ \ km \ s}}^{-1}$), and \MEASUREMENT{4}  additional candidate member stars with less precise velocity measurements.} Table \ref{table:keck_members} shows the basic properties of \CYTthree{the confirmed and candidate member} stars. Based on their DECam photometry \CYTone{and the best-fit \texttt{ugali} isochrone for Leo VI (see Figure \ref{Figure:ugali}),  we identify} these 13 Leo VI member stars as 12 Red Giant Branch (RGB) stars and 1 Blue Horizontal Branch (BHB) star.

We obtain the systemic velocity, $v_{\rm hel}$, and velocity dispersion of the system, $\sigma_v$, from the line-of-sight velocity measurements of the 9 confirmed member stars using a single two-parameter fit as described in \cite{Walker:2006}.  For the model fit, we apply a uniform prior on the systemic velocity within a range of  \MEASUREMENT{164.6  $<v_{\rm hel} <$ 178.6  km~s$^{-1}$}, based on the maximum and minimum range of velocities found in the member stars, \CYTthree{and a log-uniform prior on the velocity dispersion within a range of  $-1  <\log_{10}{(\sigma_v/{\rm{ km \ s}}^{-1})}<1$. Using \texttt{emcee} to obtain the marginalized posterior distribution (shown in Figure \ref{Figure:Keckcorner_vel}), we find that the systemic velocity of Leo VI is given by $v_{\rm hel} = \vlosval$ km~s$^{-1}$, while the velocity dispersion is given by $\sigma_v = \vdispval$ km~s$^{-1}$. We determine the values of the parameters and their uncertainties from the peak and the highest density region containing 68\% of the posterior \citep{Hyndman:1996}.} The middle subplot of Figure \ref{Figure:Leo6Keck} shows the best-fit velocity dispersion model overlaid on the histogram of the measured radial velocity of the Leo VI stars. \CYTthree{We find that the best-fit velocity dispersion remains consistent within uncertainties if we instead assume a uniform prior from $0 < \sigma_v < 10$ km~s$^{-1}$ (Figure \ref{Figure:Keckcorner_vel}).}

\begin{figure}[ht]
\centering
\includegraphics[width=0.95\linewidth]{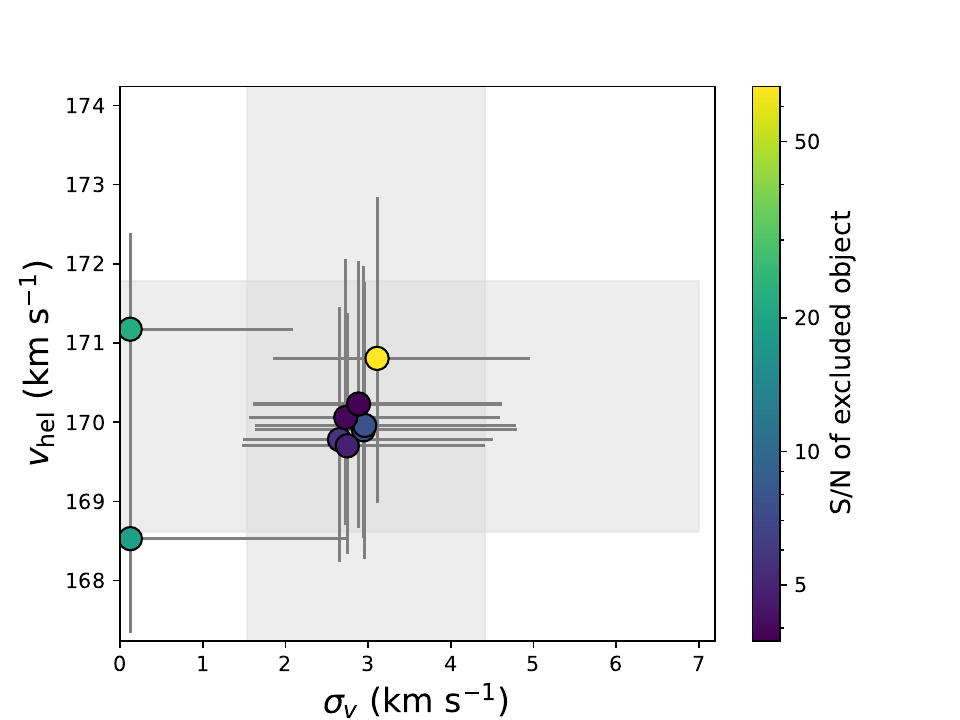}
%\vspace{-3cm}
\caption{Jackknife test for the radial velocity parameters of Leo VI. The grey regions represent  68\% percentile uncertainty of the parameters obtained from the full sample. Each point represents the parameters obtained using a subsample that has excluded one star. \CYTthree{ The velocity dispersion measurement of the system becomes unresolved when we remove either one of the two very high S/N stars (S/N$>$15) with a large difference in line-of-sight velocity between each other ($v_{\rm hel} = 164.6 \pm 1.6$, $174.0 \pm 1.7 $ km~s$^{-1}$.) \label{Figure:Jackknife} }}

\end{figure}

\begin{figure*}[ht]
\centering
\includegraphics[width=\linewidth]{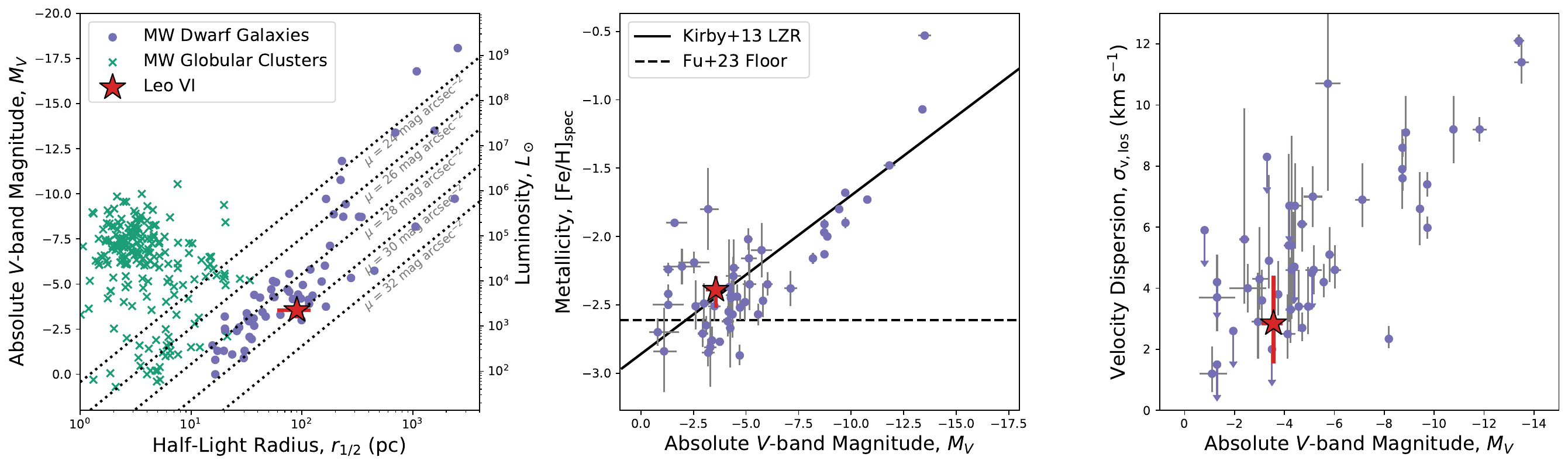}
%\vspace{-3cm}
\caption{
{\it Left}: The absolute $V$-band magnitude, $M_V$, vs.\ the azimuthally-averaged physical half-light radius, $R_{1/2}$, of Leo VI and the population of known Local Group globular clusters and dwarf galaxies. Leo VI's morphological properties are consistent with those of a dwarf galaxy.
{\it Middle}: The  systemic metallicity, $\mbox{[Fe/H]} $, vs.\  the absolute $V$-band magnitude of Leo VI and the population of Local Group dwarf galaxies. \CYTone{The overlaid black line represents the $V$-band absolute Luminosity-Metallicity Relation (LZR) found in \cite{Kirby:2013}, while the dashed black line represents the metallicity floor of \mbox{[Fe/H]} $= -2.61$ proposed for faint systems \citep{Fu:2023}}
{\it Right}: The line-of-sight velocity dispersion, $\sigma_{\rm v, los}$, vs.\ absolute $V$-band magnitude of the population of Local Group dwarf galaxies. }
\label{Figure:Population}
\end{figure*}

To further test the robustness of the systemic velocity and velocity dispersion measurements from outlier measurements, we performed a jackknife test where we recalculated the parameters for a subsample of stars where one star is removed from the total population at a time. \CYTthree{As seen in Figure \ref{Figure:Jackknife}, we found that the velocity measurements are mostly dominated by the three highest S/N stars. If we remove either of the two very high S/N stars, Gaia DR3 3993822949622182272 ($v_{\rm hel} = 164.6 \pm 1.6 $ km~s$^{-1}$)  or Gaia DR3 3993823052701408128 ($v_{\rm hel} = 174.0 \pm 1.7 $ km~s$^{-1}$), which show a large difference in line-of-sight velocity between each other, our velocity dispersion measurements become unresolved (i.e. consistent with a zero velocity dispersion measurement). For stars with ${\rm S/N} <15$, the values of parameters obtained from the jackknife test are all within $1\sigma$ of the parameters obtained from the full sample. However, since we only have single-epoch radial velocity measurements, we cannot identify any of the Leo VI member stars as unresolved binary systems that can inflate the system's velocity dispersion. Some studies suggest the binary fraction of UFD could be as high as $50\%$ \citep{Minor:2019, Arroyo:2023}. Furthermore, some of the stars could also be a non-member interloper star. }

%weighted average of five high-probability members (circled in red),

If we assume that Leo VI is a dispersion-supported system in dynamical equilibrium, we can use the estimator from \citet{Wolf:2010}  \CYTone{with the  velocity dispersion measurment of Leo VI to calculate its} enclosed mass,
\begin{equation} \label{eqn:Wolf}
    M_{1/2} = 930 M_\odot\left(\frac{\sigma_{\rm v,los}}{\rm{km \ s} ^{-1}}\right)^2\left(\frac{R_{1/2}}{\rm{pc} }\right).
\end{equation}
We estimate that the dynamical mass of the system within the half-light radius is $\dmmassval$ $M_\odot$. Assuming that the luminosity at half-light radius is given by $L_{1/2} = 0.5 L_{V}$, we obtain that the mass-to-light ratio at the half-light radius of Leo VI is given by $\mlratioval M_\odot/L_\odot$.

However, the high ellipticity of Leo VI may indicate that the system might be tidally disturbed by the MW, which would make the assumption of dynamical equilibrium invalid. We will further discuss the possibility of Leo VI being a tidally disrupted system in Section \ref{sec:tidal}.

\subsection{Metallicity and Metallicity Dispersion} \label{sec:metallicity}

In addition to the line-of-sight velocities, \texttt{DMOST} also measures the equivalent widths (EWs) of the \CYTone{infrared Ca II triplet} (CaT) lines. In this analysis, we model the CaT lines of \CYTthree{high S/N stars} (${\rm S/N} > 15$) with a combination of Gaussian and Lorentzian models, while \CYTthree{we model the rest of the stars (${\rm S/N} < 15$)}  with a single Gaussian model. \CYTone{We note that our EW measurements are only valid for RGB stars, so we exclude the EW CaT measurement for the BHB star}. \CYTthree{We also excluded EW CaT measurement for RGB stars with high line-of-sight velocity errors ($\epsilon_{v_{\rm hel}} > 10 {\rm{ \ km \ s}}^{-1}$).}

We derived $\mbox{[Fe/H]}$ metallicities from the EWs using the calibration relation from \cite{Carrera:2013}. \CYTtwo{The calibration relation requires the absolute $V$-band magnitude for each star, which we obtain using their $g$ and $r$ band magnitudes from the follow-up DECam catalog. We then convert the $g$ and $r$ magnitudes into relative $V$-band magnitude using transformation relations from \cite{DES:2021} and subtract the relative magnitudes with distance modulus from the \texttt{ugali} fit (\CYTone{Table \ref{table:Leo6_param}}) to obtain the absolute $V$-band magnitudes. }

We derive the systemic metallicity and metallicity dispersion of Leo VI using \texttt{emcee} with a similar two-parameter fit as described for the systemic velocity, \CYTthree{using only metallicity measurements for  6  member stars with S/N $>5$}. We apply a uniform prior on the spectroscopic metallicity of the system within the range of  $-4 <[\rm Fe/H]_{\rm spec}<$ 0  \CYTthree{ and a log-uniform prior on the metallicity dispersion at a range of  $-1.5 <\log_{10}{(\sigma_{[\rm Fe/H]})}< 0.5$.}  We find that the \CYTone{systemic} spectroscopic metallicity of Leo VI is $[\rm Fe/H]_{\rm spec} =\fehval$ and a metallicity dispersion at $\sigma_{[\rm Fe/H]} = \fehdispval$ (see Figure \ref{Figure:Keckcorner_feh}). \CYTthree{Similar with the velocity dispersion measurement, we repeated the analysis with uniform priors on the metallicity dispersion within a range of $0 <\sigma_{[\rm Fe/H]}<$ 3 and find that the metallicity and metallicity dispersion are within 1$\sigma$ of the original values (Figure \ref{Figure:Keckcorner_feh}). }  

\CYTtwo{As shown in Figure \ref{Figure:Population}, we find the size-luminosity, luminosity-metallicity, and luminosity-velocity dispersion relation of Leo VI to be consistent with other MW satellite galaxies.}

\Needspace{5\baselineskip}
\section{Discussion} \label{sec:discussion}
There are several unique properties of Leo VI that make its discovery particularly interesting, \CYTone{namely its highly elliptical shape and proximity to other MW satellites}. We first discuss the orbital properties of Leo VI derived by combining Keck radial velocity measurements with \textit{Gaia} proper motion measurements in Section \ref{sec:orbit}. We then discuss whether \CYTone{Leo VI's elliptical shape might indicate that it} is undergoing tidal disruption in Section \ref{sec:tidal}. In Section \ref{sec:infall}, we consider the possibility of Leo VI being part of a group infall scenario due to its proximity to other satellite galaxies in the constellations of Leo and Crater.

\begin{figure*}[ht]
\centering
\includegraphics[width=\linewidth]{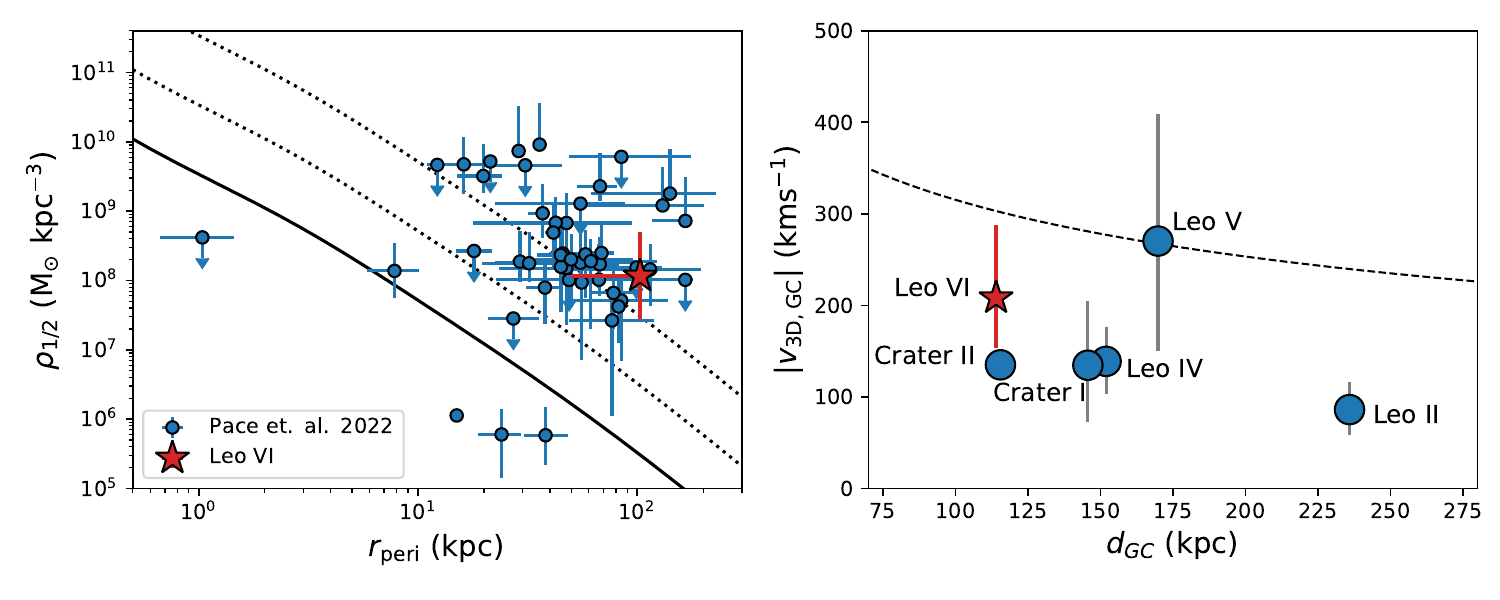}
\caption{{\it Left}:  The average density within a half-light radius, $\rho_{1/2}$, vs.\ distance of pericenter, $r_{\rm peri}$, of  Leo VI and other MW satellites \citep{Pace:2022}. The solid black line represent twice the enclosed MW \CYTone{density as a function of radius, while the dashed lines represent 20 and 200 times the enclosed MW density. If a satellite is located below the solid line, its half-light radius is estimated to be bigger than its Jacobi (tidal) radius and it} is likely to be tidally disrupting \citep{Pace:2022}.
{\it Right}: The magnitude of the 3D Galactocentric velocity, $|V_{\rm 3D, GC}|$,  vs.\ the Galactocentric distance, $d_{\rm GC}$, of Leo VI and other nearby satellite galaxies. The dashed lines represented the escape velocity of the potentials from the \texttt{gala}'s \texttt{MilkyWayPotential} potential. \CYTone{Leo VI is likely a bound system, and thus it is likely to have reached its orbital pericenter at least once in the past.}}
\label{Figure:MWPotential}
\end{figure*}

\Needspace{5\baselineskip}
\subsection{Proper Motion and Orbit of Leo VI}
\label{sec:orbit}
To obtain three-dimensional velocity information of the system, we cross-match our DECam-based stellar catalog with the \textit{Gaia} \CYTone{DR3} \citep{Gaia:2023}.  As shown in the right subplot of Figure \ref{Figure:Leo6Keck}, all 3 spectroscopically confirmed member stars of Leo VI  which also have \textit{Gaia} proper motion measurements have proper motions clustered near $\mu_\delta, \mu_\alpha \cos(\delta) \sim (0,0)$\, mas/yr, further confirming that it is a gravitationally-bound system that is located far from the MW. Using \texttt{emcee} to fit a Gaussian mixture model and \CYTone{taking into account} for the correlations in $\mu_{\alpha * }$ and $\mu_{\delta}$, we find that the systemic proper motion of Leo VI is $\mu_{\alpha * } = \pmraval$  mas yr$^{-1}$ and $\mu_{\delta} = \pmdecval$  mas yr$^{-1}$.  

To determine the orbit of Leo VI, we integrated 1,000 realizations of its orbit using the \texttt{gala} galactic dynamics package \citep{Price-Whelan:2017}. We obtain a sample of the possible current 6-D positions and velocities ($\alpha_{J2000}$, $\delta_{J2000}$, $D_\odot$, $\mu_{\alpha * }$, $\mu_{\delta}$, $v_{\rm hel}$) of the system by sampling from the Gaussian error distribution of the observed position and velocity parameters (see Table \ref{table:Leo6_param}), \CYTone{and convert it to the \texttt{astropy} v4.0  Galactrocentric Frame \citep{2013A&A...558A..33A}.} We then rewind Leo VI’s orbit back in time for 10 Gyrs in the presence of \texttt{gala}’s \texttt{MilkyWayPotential} model \citep{Bovy:2015} and recorded \texttt{gala}'s estimate of the orbital parameters of Leo VI. We find an orbital apocenter of $\apoval$ kpc, while the pericenter is $\perival$ kpc and the orbital eccentricity is  $\orbeccval$. \CYTthree{Following the velocity and metallicity dispersion analyses, we obtain the best-fit values and uncertainties of the parameters from the peak of the distribution and the highest density region containing 68\% of the posterior, respectively.}
\CYTthree{At a heliocentric distance of D = $\distval$\, kpc, proper motion uncertainties of 0.2  mas/yr corresponds to velocities uncertainties  of $\sim$100  km s$^{-1}$, introducing large uncertainties into our orbit estimates}.

Moreover, \CYTtwo{the z-component of the specific angular momentum, $L_z$, and the specific orbital energy} of the system, $E$, are given by $L_z = \orbLzval$ kpc$^2$ Myr$^{-1}$ and $E = \orbEval$ kpc$^2$ Myr$^{-2}$, respectively. \CYTone{We further find that \orbEpctval\% of realizations of Leo VI's orbit is bounded ($E<0$ kpc$^2$ Myr$^{-2}$) and \orbLzpctval\% of realizations yield a prograde orbit.}

\begin{figure*}[ht]
\centering
\includegraphics[width=0.9\linewidth]{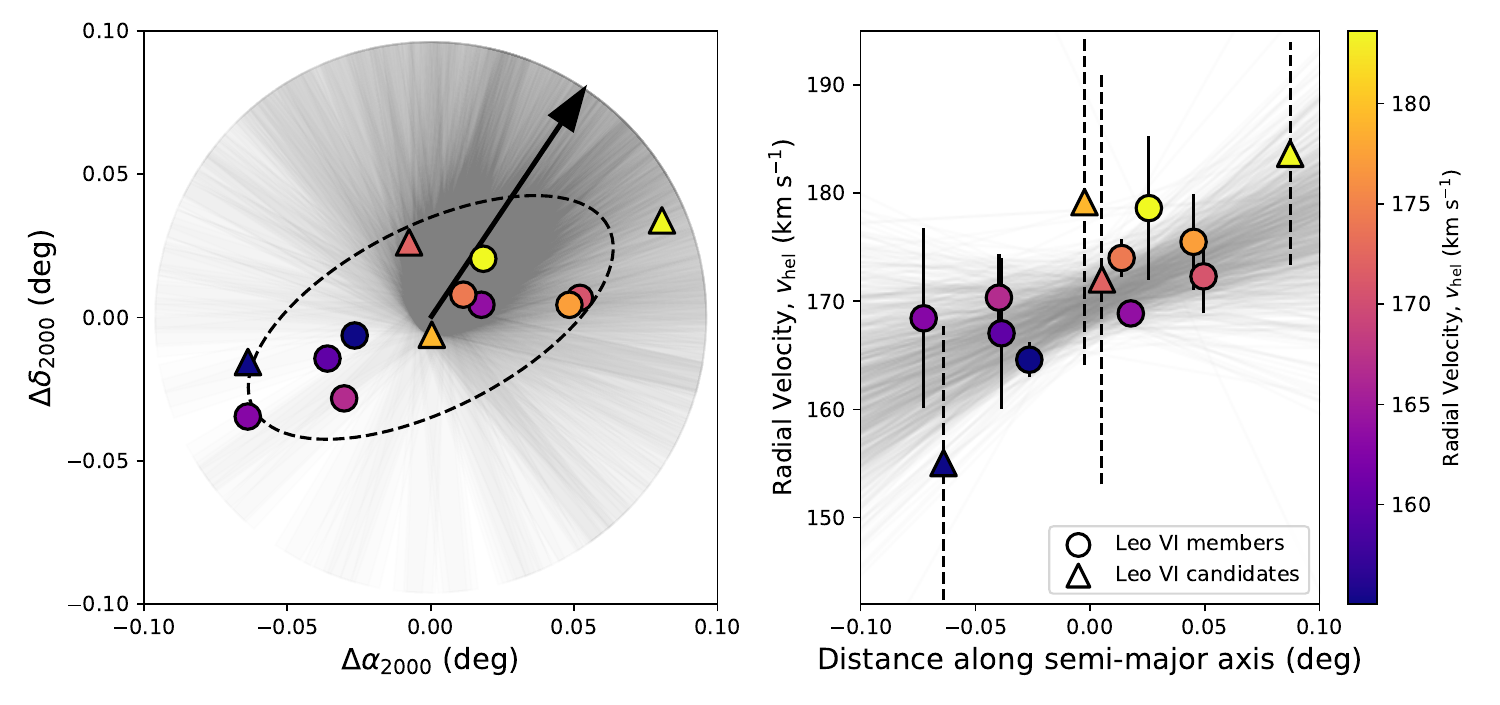}
%\vspace{-3cm}
\caption{
{\it Left}:  Spatial distribution of the spectroscopically confirmed Leo VI \CYTthree{member stars (circles) and candidate member stars (triangles)} with points colored by the measured line-of-sight (radial) velocity. The dashed ellipse shows the half-light radius of Leo VI. The direction of the median systemic proper motion (after correcting for solar reflex motion), computed from the three spectroscopically confirmed members with \textit{Gaia} proper motion measurements, are shown with the black arrow, while the faint gray lines represent the 1,000 Monte Carlo samplings from the proper motion uncertainties. The direction of proper motion aligns with the semi-major axis of Leo VI.  
{\it Right}: The line-of-sight (radial) velocity of Leo VI \CYTthree{member stars (circles) and candidate member stars (triangles with dashed error bars)} as a function of distance along its semi-major axis. The gray lines represent the  Monte Carlo samplings used to measure the radial velocity gradient of the system along the semi-major axis. \CYTone{Due to the low number statistics and high uncertainty of the radial velocity measurements, we are unable to conclusively determine the presence/absence of a velocity gradient.}
}
\label{Figure:Tides}
\end{figure*}

\subsection{Possible Tidal Disruption}
\label{sec:tidal}

Due to the high ellipticity of Leo VI, coupled with the fact that the direction of systemic proper motion aligns with the semi-major axis (see left panel of Figure \ref{Figure:Tides}), we discuss the possibility of the system being tidally disturbed by the MW. Common dynamical mass estimators for UFDs (such as \citealt{Wolf:2010})  are based on the assumption that the system is in dynamical equilibrium, which doesn't apply to disrupting systems. Therefore, determining whether the system is in dynamical equilibrium or has been tidally disturbed has important implications for measurements of its dark matter content.

To assess whether the system is experiencing tidal disruption, we follow  the methodology of  \cite{Pace:2022}  and compare \CYTtwo{the average density of Leo VI within its half-light radius}, $\rho_{1/2} = \rhoLeoval M_\odot $ kpc$^{-3}$,  to twice the average MW density at its orbital pericenter {$2\rho_{MW}(r=r_{\rm peri})= \rhoMWval M_\odot $ kpc$^{-3}$. \CYTone{If we assume a flat rotation curve for the MW and that Leo VI has an circular orbit, this comparison is equivalent to comparing the half-light radius of the system with its Jacobi (or tidal) radius where beyond it the  tidal forces exceed the systems own gravitational force.} As illustrated in Figure \ref{Figure:MWPotential}, the average density Leo VI is much higher than twice the average MW density. \CYTone{Therefore, Leo VI's Jacobi radius is likely larger than its half-light radius even when it is at its orbital pericenter, and that it is unlikely that the system is tidally disrupting.} \CYTtwo{This conclusion also holds even when using 1$\sigma$ lower bound for both the average density of Leo VI and pericenter distance.} \CYTone{However, we note that this approximation does not hold for all systems as the FIRE simulation has found the presences of high density satellites galaxies which are still tidally disrupting \citep{Shipp:2023}}.  In addition, Figure  \ref{Figure:MWPotential}  shows the measured 3D velocity of Leo VI with other nearby satellites compared to the local escape velocity of   \texttt{gala}'s \texttt{MilkyWayPotential} potential. It is likely that Leo VI is a bound system and that it is not on first infall, allowing the possibility of the system to reach its pericenter (at $\perival$\,kpc) in the past.

Another signature of a tidally disrupting system is the presence of a velocity gradient in its member stars. \CYTone{Figure \ref{Figure:Tides} illustrates the radial velocity of the spectroscopically confirmed member stars of Leo VI as a function of its on-sky position and distance along the semi-major axis. To measure the velocity gradient of the system, we use a linear model to fit the radial velocity of the \CYTthree{9} Leo VI members stars as a function of its distance along the semi-major axis while taking account of intrinsic scatter.} \CYTthree{Using \texttt{emcee} to sample the posterior of the linear model, we find a large best-fit velocity gradient with large uncertainties, $\velgradval $ km s$^{-1}$/deg ($\velgraduval $ km s$^{-1}$/kpc). 
When performing a Bayesian Information Criterion test, we find that the non-zero velocity gradient model is mildly preferred over the zero velocity gradient model at $\Delta_{\rm BIC,01} = \velgradbic$. However, we note that as with the velocity dispersion analysis, the velocity gradient analysis is dominated by the three highest S/N stars, and that removing any of these stars from the sample has a large impact on the velocity gradient. For example, if we remove the member star   \textit{Gaia} DR3 3993822949622182272 from our sample, the velocity gradient drops to $\velgradjackval $ km s$^{-1}$/deg ($\velgradjackuval $ km s$^{-1}$/kpc). Therefore, we caution that the apparent velocity gradient might be a product of small number statistics and that more measurements are needed.}

\begin{figure*}[ht]
\centering
\includegraphics[width=\linewidth]{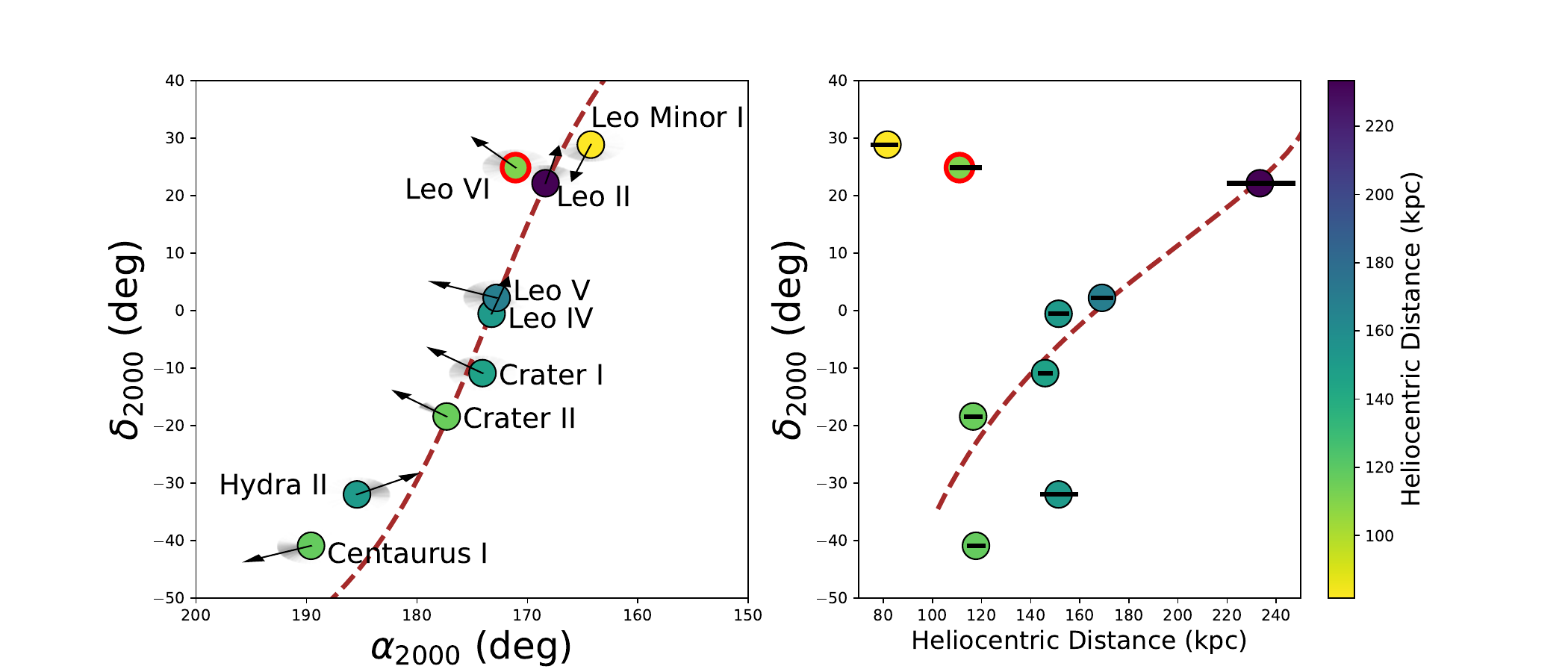}
%\vspace{-3cm}
\caption{
{\it Left}:  Spatial distribution of Leo VI and other potential members of the Crater-Leo infall group. The markers are colored based on the heliocentric distance of the system. \CYTone{The black arrows represent the  direction of the median systemic proper motion of the system (after correcting for solar reflex motion), while the faint grey lines represent the 1,000 Monte Carlo samplings from the proper motion uncertainties. The proper motion  for the members of the Crater-Leo infall group were obtained from \cite{Julio:2024}, where we have use their Hubble Space Telescope measurements for Leo IV and Leo V.} The brown line represents a great circle with a pole centered at (83.2$^\circ$, -11.8$^\circ$) that passes close to the members of the infall group \citep{Torrealba:2016}.
{\it Right}:  The Heliocentric distance of  Leo VI  and the nearby satellites as a function of the declination, $\delta_{2000}$. The brown line shows the orbit of the proposed Leo-Crater group from \cite{Torrealba:2016}. The plots indicate that the heliocentric distances of Leo VI and Leo Minor I are inconsistent with the distance gradient followed by the members of the Crater-Leo group.
% {\it Right}: The same subplot as the middle subplot but in the  X-Z projection. 
  }
\label{Figure:LeoFriends}
\end{figure*}

\CYTtwo{Moreover, both observations of MW satellites \citep{Pace:2022} and N-body simulations \citep{Munoz:2008}} has found no strong correlation between the ellipticity of the system and whether it is tidal disrupting.  \CYTone{\cite{Pace:2022} also observed that the } proper motion vector of \CYTone{highly elliptical} UFDs often aligns with its semi-major axis even for non-disrupting systems with large pericenters.

%and a distribution ranging from -32 to 172 km s$^{-1}$/deg  (corresponding to 95\% confidence interval)

%While it is often assumed that UFDs at large galactocentric distances are not affected by the tides of the MW, spectroscopic observations of Antlia II and Crater II (which are located at a distance of $>$ 100 kpc from the MW) have revealed that these systems are experiencing tidal disruption \citep{Ji:2021}. 

\begin{figure*}[ht]
\centering
\includegraphics[width=\linewidth]{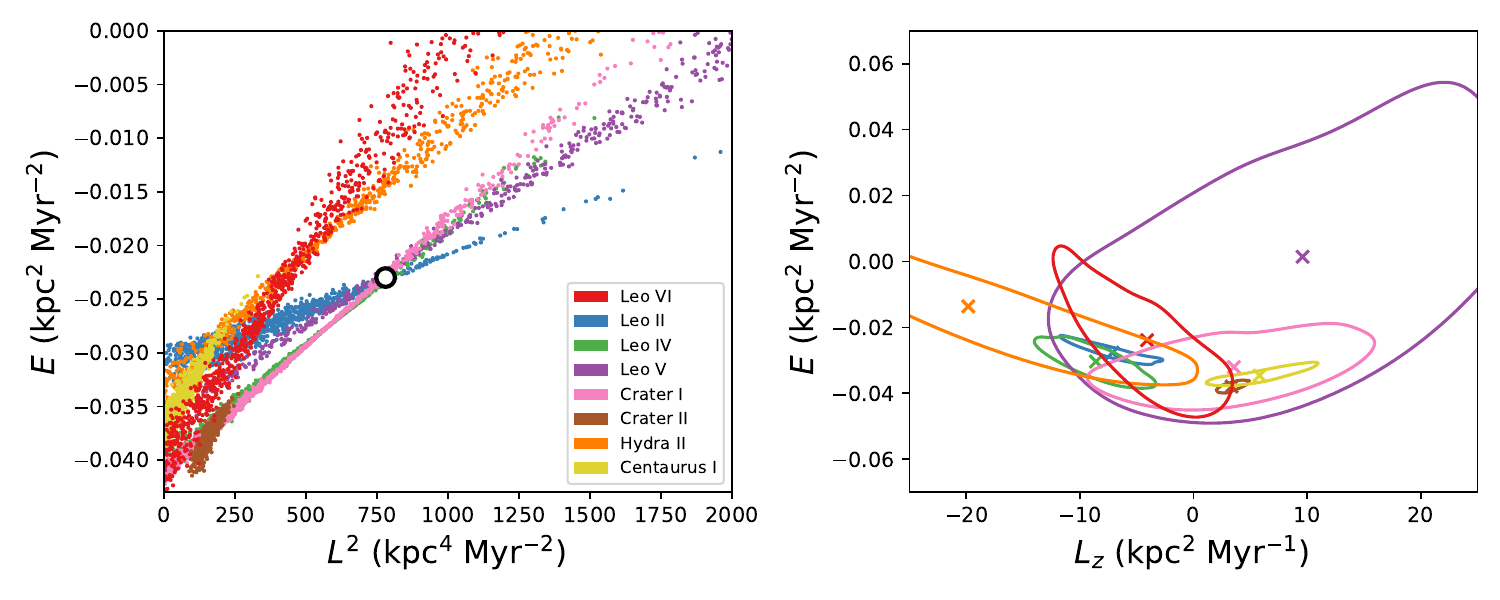}
%\vspace{-3cm}
\caption{
{\it Left}:  The specific energy, $E$, vs.\ the square of the specific 
angular momentum, $L^2$, of Leo VI and the members of the Crater-Leo infall group. We find that the $E-L^2$ distribution of  Leo II, Leo IV, Leo V, and Crater all intersect at $(L^2, E ) \sim$ (780 kpc$^4$ Myr$^{-2}$ , $-0.023$ kpc$^2$ Myr$^{-2}$), suggesting a common origin \citep{Pawlowski:2021}.  The $E-L^2$ distribution of Leo VI is inconsistent with the other infall group members, suggesting that it may not be part of the group.
\CYTone{{\it Right}: The specific energy, $E$, vs.\ the z-component of the specific angular momentum, $L_z$,  of Leo VI and the other satellites. The cross markers represent the median $E-L_z$ values of the satellites, while the contour lines represent the 1$\sigma$ uncertainty. $E-L_z$ distribution of Leo VI is  consistent with all the other members of the Crater-Leo infall group, albeit with large uncertainties.}}
\label{Figure:LeoCrater2}
\end{figure*}

\Needspace{5\baselineskip}
\subsection{Group Infall scenario}
\label{sec:infall}

The on-sky location of Leo VI is close to several other known distant MW satellite galaxies that are already found in the constellations Leo and Leo Minor. Due to their similar on-sky positions and radial velocity, it has been suggested that the MW satellite galaxies in Leo  (Leo II, Leo IV, Leo V), the star cluster Crater/Laevens 1, and the dwarf galaxy Crater II might have been accreted into the  MW through a group infall
scenario \citep{Torrealba:2016, Pawlowski:2021, Julio:2024}. \CYTone{Moreover, simulations from \cite{Li:2008} have shown that about one-third of subhaloes are accreated into the MW through infall groups. } In this section, we discuss the possibility of Leo VI \CYTone{and other close-by systems such  Leo Minor I \citep{Cerny:2023c}, Hydra II \citep{Martin:2015} and Centaurus I \citep{Mau:2020}} being members of the proposed Leo-Crater infall group. 

\cite{Torrealba:2016} found that the member systems of the Leo-Crater infall group are all located close to the great circle with the pole at ($\alpha, \delta$) = (83.2$^\circ$, $-$11.8$^\circ$) and forms a consistent heliocentric distance gradient as a function of  their declination. Figure \ref{Figure:LeoFriends} shows the heliocentric distance of Leo VI and the other members of the infall group as a function of their spatial distribution. From the figure, we can see that both Leo VI and especially Leo Minor I are too close to follow the heliocentric distance gradient exhibited by the other members of the infall group. \CYTtwo{While the distances of Hydra II and Centaurus I are more consistent with the other members of the group, their distances are not consistent with the distance trend from \cite{Torrealba:2016}}.

We expect that satellites that are accreted into the MW via an infall group to share similar values of total energy and angular momentum \citep{LydenBell:1995}. Using \texttt{gala}’s \texttt{MilkyWayPotential} model and the velocity measurements of the Leo-Crater group collected from \cite{Julio:2024}, we calculated the  specific angular momentum  and specific energy distribution of the member satellites of the Leo-Crater group. \CYTone{For Hydra II and  Centaurus I, we use proper motion and radial  velocity measurements from \cite{,Kirby:2015, Pace:2022} and \cite{Heiger:2024}, respectively. We exclude Leo Minor I from the following analysis due to the lack of radial velocity measurements. } As shown in Figure \ref{Figure:LeoCrater2},  the distribution of specific energy, $E$, and the square of the specific angular momentum squared, $L^2$, of 4 satellites (Leo II, Leo IV, Leo V and Crater) all intersect each other at $(L^2, E ) \sim$ \MEASUREMENT{(780 kpc$^4$ Myr$^{-2}$ , -0.023 kpc$^2$ Myr$^{-2}$}) suggesting a common origin \citep{Pawlowski:2021}. We find that the $E-L^2$  distribution of Crater II, Leo VI, Hydra II and Centaurus I is inconsistent with the other satellites of the group. However, as shown in Figure~\ref{Figure:LeoCrater2}, \CYTone{the specific energy, $E$, and z-component of specific angular momentum $E-L_z$ distribution of Leo VI is consistent with all the other members of the proposed group (including Crater II). Although this is mostly due to the large uncertainties in the proper motion measurement of Leo VI as Crater II has a inconsistent $E-L_z$ distribution with some of the other members of the infall group such as Leo II and Leo IV.}

\CYTone{Systems that are part of an infall group are also expected to have similar orbital poles or direction of angular momentum.  \cite{Julio:2024} found that 4 systems in the proposed Leo-Crater Group (Leo II, Leo IV, Leo V and Crater) have orbital poles that all  intersect with each other, while position of the orbital poles of Crater II doesn't not match the other members of the system. As shown in Figure \ref{Figure:OrbitalPoles}, the position of the orbital pole of Leo VI is consistent with the group orbital pole  of the Leo-Crater infall group at $(\alpha_{J2000},\delta_{J2000}) \sim (210^{\circ},-24^{\circ})$ while Hydra II  also has orbital pole that is somewhat close by ($\sim 6 \deg$ away). As with the energy-angular momentum distributions, we find that position  orbital pole of Centuarus I is inconsistent with the other members of the proposed group.}

\CYTone{While Leo VI shares a similar on-sky position,  orbital pole location and $E-L_z$ distribution with the members of the Leo-Crater  infall group, due to its relatively close distance and inconsistent  $E-L^2$ distribution with the other members of the group, we determine that it is unlikely for Leo VI being a part of the the proposed infall group. However, more precise measurements of the proper motion of Leo VI are needed to more definitely determine the its membership in the Leo-Crater group. Similarly, we also did not find convincing evidence that Leo Minor I, Hydrus II and  Centuarus I are part Leo-Crater infall group due to their inconsistent heliocentric distance, $E-L^2$ distribution and orbital pole location with the infall group.  }

%In particular, Leo VI is at a similar distance as the nearby UFD satellite Crater II at 120 kpc \citep{torrealba16}. Therefore, orbital information of  Leo VI obtained derived from radial velocities from future follow-up observations can provide evidence to verify the galaxy's membership in the proposed Leo-Crater infall group and thus provide further evidence of hierarchical galaxy formation at the smallest scales.

\begin{figure*}[ht]
\centering
\includegraphics[width=\linewidth]{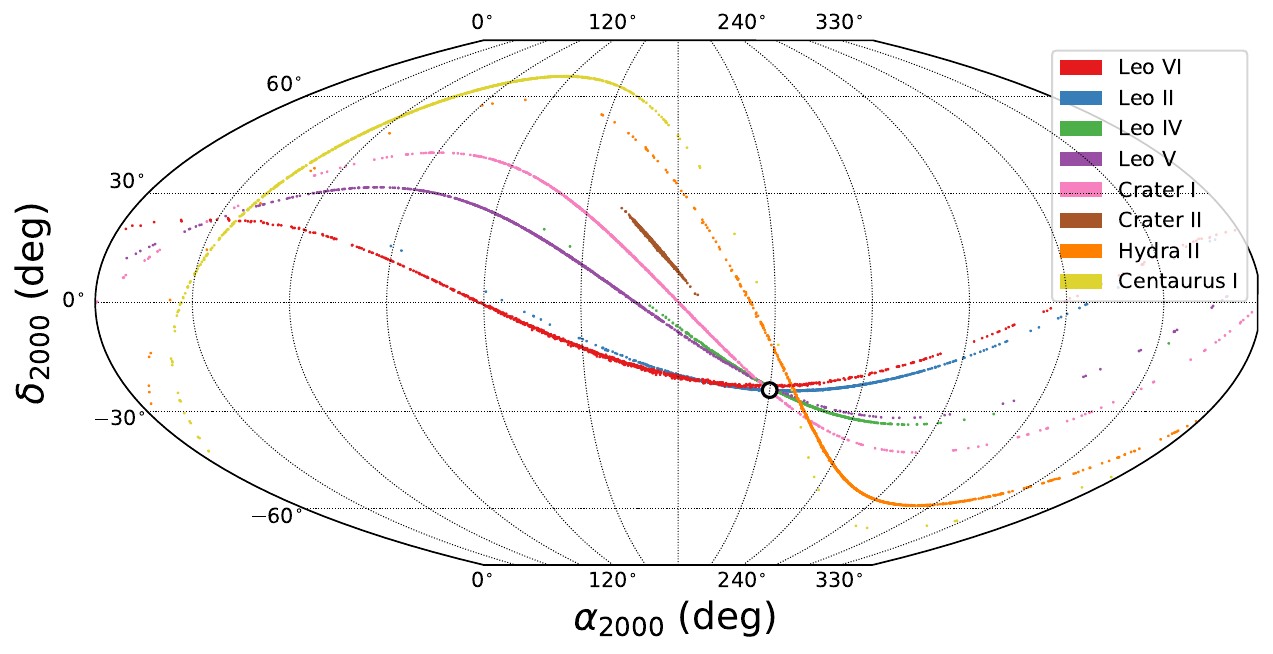}
%\vspace{-3cm}
\caption{
\CYTone{Distribution of the possible orbital poles of Leo VI and the other possible members of the Crater-Leo infall group. The position of the orbital pole of Leo VI is consistent with the possible group orbital  pole of the Crater-Leo group (Leo II, Leo IV, Leo V and Crater) at $(\alpha_{J2000},\delta_{J2000}) \sim (210^{\circ},-24^{\circ})$, while the position of the orbital pole of Crater II is inconsistent with the  group orbital  pole.}}
\label{Figure:OrbitalPoles}
\end{figure*}

%
% \textbf{}
\Needspace{5\baselineskip}
\subsection{Astrophysical J-factor}
\CYTone{As mentioned in Section \ref{sec:intro}, MW satellite UFDs are also useful targets for  searches for dark matter annihilation or decay products \citep{Ackermann:2015zua, Geringer-Sameth:2015, McDaniel:2023, Boddy:2024}. The astrophysical component that governs the dark matter annihilation and decay fluxes are referred to as the J-factor and D-factor, respectively. Both factors depend on the dark matter density of the system along the line of sight such that $J(\theta) = \iint_{\rm los} \rho_{\rm DM}^2  \rm{d} l \rm{d}\Omega $ and  $D(\theta) = \iint_{\rm los} \rho_{\rm DM} \rm{d} l  \rm{d}\Omega $, where $\rho_{\rm DM}$ is the dark matter density, $\rm{d}l$ is the line-of-sight direction and $\rm{d}\Omega$ is the solid angle with radius $\theta$}.

\CYTtwo{We use the framework developed by \cite{Bonnivard2015MNRAS.453..849B, GeringerSameth2015ApJ...801...74G, Pace:2019}} to calculate the J-factor and D-factor of Leo VI. We model the system with a stellar component parameterized by a Plummer distribution \citep{Plummer:1911} and dark matter component with a Navarro–Frenk–White profile \citep{Navarro:1997} while assuming a constant stellar anisotropy with radius. We then fit the dark matter profile by modelling the velocity dispersion of the member stars using the  spherical Jeans equations and compared them with the measured velocity dispersion from the spectroscopic observations.

For the J-factor of Leo~VI \CYTthree{obtained using measurements from the 9 member stars}, we find \MEASUREMENT{$\log_{10}{J(\theta)}=17.0_{-1.0}^{+0.8}$, $17.2_{-1.0}^{+0.8}$,  $17.3_{-1.0}^{+0.9}$, and $17.3_{-1.0}^{+0.9}$ }  for solid angles of $\theta=0.1^{\circ}, 0.2^{\circ}, 0.5^{\circ}, 1^{\circ}$ in logarithmic units of ${\rm GeV^{2}~cm^{-5}}$. These J-factor measurements are consistent with estimates of \MEASUREMENT{$\log_{10}(J(0.5^{\circ})) = 16.9_{-1.2}^{+0.8}$}, obtained from the empirical scaling relation derived in \cite{Pace:2019},
\begin{footnotesize}
\begin{equation}
    \frac{J(0.5^{\circ})}{\rm GeV^{2}~cm^{-5}} = 10^{17.87}\left(\frac{\sigma_{\rm v,los}}{\rm {5 ~ km s}^{-1}}\right)^4 \left(\frac{D_\odot}{\rm {100~kpc}}\right)^{-2} \left(\frac{R_{1/2}}{\rm {100 ~pc}}\right)^{-1} .
\end{equation}
\end{footnotesize}

For the D-factor of Leo~VI, we find \MEASUREMENT{$\log_{10}{D(\theta)}=16.6\pm0.5$, $17.0_{-0.6}^{+0.5}$, and $17.5\pm_{-0.7}^{+0.6}$},   for solid angles of $\theta=0.1^{\circ}, 0.2^{\circ}, 0.5^{\circ}$ in logarithmic units of ${\rm GeV~cm^{-2}}$. As the velocity dispersion has a tail to zero velocity dispersion \CYTone{(see Figure \ref{Figure:Keckcorner_vel})}, the J-factor and D-factor also have similar tails. We have applied a $V_{\rm max} > 1~{\rm km~s^{-1}}$ cut to remove  the tail and note that without this change the J-factor and D-factor decrease by roughly 0.05 dex.

While Leo VI does not have a large J-factor compared to other MW UFDs, \CYTone{for example Wilman 1 and Ursa Major II have J-factors of  $\log_{10}J(0.5^{\circ}) = 19.53_{-0.50}^{+0.50}$  and $19.44_{-0.39}^{+0.41}$   GeV$^2$cm$^{-5}$, respectively \citep{Pace:2019}}, it could be included in a stacked analysis with other UFDs.

\Needspace{5\baselineskip}
\section{Summary}
\label{sec:summary}
\CYTtwo{We have presented the DELVE EDR3 data set and used it to discover the MW satellite Leo VI.} Using the \texttt{ugali} maximum likelihood-fit of the systems morphology and color-magnitude diagram, we find that Leo VI is an old (\CYTone{$\tau>$\ageval \ Gyrs}), metal-poor ($\mbox{[Fe/H]}_{\rm spec}  = \fehval$), stellar system with low luminosity ($M_V =\mvval $ mag), large size ($R_{1/2} = \rhalfval $ pc), elliptical shape ($\epsilon = \ellipval$) and a large heliocentric distance ($D = \distval$\, kpc). By obtaining Keck/DEIMOS spectroscopy of the possible member stars of Leo VI, we are able to find \CYTthree{ 9 member stars  and 4 candidate member} stars for the system (with line-of-sight velocities of  155 kms$^{-1}$ $\leq v_{\rm hel} \leq $ 185 kms$^{-1}$), and find a non-zero velocity dispersion of  $\vdispval$ km s$^{-1}$. We are also able to obtain orbital properties of Leo VI by combining proper motion measurements from \textit{Gaia} DR3 ($\mu_{\alpha * } = \pmraval$ mas yr$^{-1}$, $\mu_{\delta}=\pmdecval$ mas yr$^{-1}$) with the system’s measured radial velocity of $v_{\rm hel} = \vlosval$ km s$^{-1}$.

Since their discovery in SDSS, there has been considerable uncertainty in the classification of some ultra-faint MW satellites as either UFDs or globular clusters \citep{Simon:2019}.  As illustrated in Figure \ref{Figure:Population}, the half-light radius of Leo VI ($R_{1/2} = \rhalfval $ pc) is bigger than all known Milky Way globular clusters, likely indicating its identity as an UFD.

Compared to globular clusters, UFDs also have much higher amounts of dark matter \citep{Simon:2019}. \CYTthree{If we assume that Leo VI is a self-gravitating star cluster, we can estimate the line-of-sight velocity using Eqn. \ref{eqn:Wolf} by assuming that the total mass within the half-light radius is half of the stellar mass. We thus estimate that if Leo VI's stars are self-gravitating, they would have a velocity dispersion of $\sigma_v \sim$~0.2 km~s$^{-1}$. This is much lower than the measured dispersion of $ \vdispval$  km s$^{-1}$ 
 further supporting the dark-matter-dominated nature of the system. We preform a Bayesian Information Criterion test and find that the self-gravitating model ($\sigma_v =$ 0.2  km~s$^{-1}$) is disfavoured over the free $\sigma_v$ parameter model  at $\Delta_{\rm BIC,01} = \veldispbic$, which approximately equivalent to a Bayes factor of \veldispbayes \
 \citep[BF$_{10} \sim \exp{(\Delta_{\rm BIC,01}/2)}$;][]{Kass:1995}. }

\CYTthree{We also find a non-zero metallicity dispersion measurement of $\sigma_{[\rm Fe/H]}$ =  $0.19_{-0.11}^{+0.13}$, which suggest the presence of a dark matter halo massive enough to retain the supernove ejecta needed for multiple generations of star formation \citep{Simon:2019}.}

Despite the system's highly elliptical shape and the alignment of its proper motion vector with the semi-major axis,  we have found that Leo VI's average density within its half-light radius is much larger than average MW density at its orbital pericenter, disfavoring the idea that the system is undergoing disruption.  The system also has an on-sky location that is close to other members of the proposed Crater-Leo infall group. However,  it is improbable that Leo VI is a part of this group as its heliocentric distance and energy-angular momentum ($E-L^2$) distribution do not match the distribution found in other members.

Leo VI is the \MEASUREMENT{fourteenth} \CYTtwo{ultra-faint MW satellite found in DELVE data (\citealt{Mau:2020, Cerny:2021,Cerny:2021b,Cerny:2023c,Cerny:2023,Cerny:2023b, Cerny:2024}).} However, unlike the other DELVE satellites, it was found in the preliminary data from the DELVE EDR3 catalog, which features deeper and more accurate photometric measurements than previous DELVE releases \citep{Drlica-Wagner:2022}. Therefore, it is expected that a more comprehensive dwarf galaxy search of the upcoming full DELVE Data Release 3 catalog will yield many more new MW satellite discoveries. \cite{Manwadkar:2022} forecasted that the final DELVE-WIDE survey is expected to find $64_{-13}^{+17}$ MW satellites with $M_V < 0 $ and $R_{1/2} >$ 10 pc at $\delta_{J2000}<0$ deg,\footnote{\CYTthree{The DELVE-WIDE survey has subsequently expanded coverage to $\delta_{J2000} \lesssim 30$ deg.}} 
while only \MEASUREMENT{35 systems} have been discovered in this region so far. Furthermore, it is expected that the upcoming  Vera C. Rubin Observatory's
Legacy Survey of Space and Time \citep{Ivezic:2019} will discover hundreds of UFDs in the local volume \citep{Hargis:2014, Mutlu-Pakdil:2021, Manwadkar:2022}. Space-based telescopes such as Roman \citep{Spergel:2015} and Euclid \citep{Euclid:2022} are also expected to have the potential of finding more ultra-faint dwarf galaxies \citep{Nadler:2024}. This growing population of known UFDs in the Local Group will allow us to probe the matter power spectrum to even smaller scales and provide insight into the process of galaxy formation for the smallest galaxies.

\Needspace{5\baselineskip}
\section*{Acknowledgments}
\CYTthree{We thank the anonymous referee for the many useful comments that helped us improve this manuscript.} The follow-up DECam observations taken on June 2023 was obtained during a time trade with the Dark Energy Camera Legacy Survey (DECaLS) team. \CYTthree {We note that Leo VI was also independently identified as a possible UFD  among ${\sim}\,300$ candidates reported in an unpublished search of PS1 data by \citet{Grillmair:2018}.}

CYT was supported by the U.S. National Science Foundation (NSF) through the grants AST-2108168 and AST-2307126. WC thanks Michael Lundquist and the staff of the W.M. Keck Observatory for extensive assistance on the night of our Keck observations, and acknowledges support from a Gruber Science Fellowship at Yale University.
ABP acknowledges support from NSF grant AST-1813881. JAC-B acknowledges support from FONDECYT Regular N 1220083. C.E.M.-V. is supported by the international Gemini Observatory, a program of NSF NOIRLab, which is managed by the Association of Universities for Research in Astronomy (AURA) under a cooperative agreement with the U.S. National Science Foundation, on behalf of the Gemini partnership of Argentina, Brazil, Canada, Chile, the Republic of Korea, and the United States of America. GEM acknowledges support from the University of Toronto Arts \& Science Post-doctoral Fellowship program, the Dunlap Institute, and the Natural Sciences and Engineering Research Council of Canada (NSERC) through grant RGPIN-2022-04794. DJS acknowledges support from NSF grant AST-2205863.

The DELVE project is partially supported by Fermilab LDRD project L2019-011, the NASA Fermi Guest
Investigator Program Cycle 9 grant 91201, and the
U.S. National Science Foundation (NSF) under grants AST-2108168 and AST-2307126. This work is supported by the Fermilab Visiting Scholars Award Program from the Universities Research Association. \CYTthree{This material is also based upon work supported by the National Science Foundation Graduate Research Fellowship Program under Grant No. DGE2139841. Any opinions, findings, and conclusions or recommendations expressed in this material are those of the author(s) and do not necessarily reflect the views of the National Science Foundation. }

\CYTone{This project used data obtained with the Dark Energy Camera (DECam), which was constructed by the Dark Energy Survey (DES) collaboration. Funding for the DES Projects has been provided by the US Department of Energy, the U.S. National Science Foundation, the Ministry of Science and Education of Spain, the Science and Technology Facilities Council of the United Kingdom, the Higher Education Funding Council for England, the National Center for Supercomputing Applications at the University of Illinois at Urbana-Champaign, the Kavli Institute for Cosmological Physics at the University of Chicago, Center for Cosmology and Astro-Particle Physics at the Ohio State University, the Mitchell Institute for Fundamental Physics and Astronomy at Texas A\&M University, Financiadora de Estudos e Projetos, Fundação Carlos Chagas Filho de Amparo à Pesquisa do Estado do Rio de Janeiro, Conselho Nacional de Desenvolvimento Científico e Tecnológico and the Ministério da Ciência, Tecnologia e Inovação, the Deutsche Forschungsgemeinschaft and the Collaborating Institutions in the Dark Energy Survey.
The Collaborating Institutions are Argonne National Laboratory, the University of California at Santa Cruz, the University of Cambridge, Centro de Investigaciones Enérgeticas, Medioambientales y Tecnológicas–Madrid, the University of Chicago, University College London, the DES-Brazil Consortium, the University of Edinburgh, the Eidgenössische Technische Hochschule (ETH) Zürich, Fermi National Accelerator Laboratory, the University of Illinois at Urbana-Champaign, the Institut de Ciències de l’Espai (IEEC/CSIC), the Institut de Física d’Altes Energies, Lawrence Berkeley National Laboratory, the Ludwig-Maximilians Universität München and the associated Excellence Cluster Universe, the University of Michigan, NSF NOIRLab, the University of Nottingham, the Ohio State University, the OzDES Membership Consortium, the University of Pennsylvania, the University of Portsmouth, SLAC National Accelerator Laboratory, Stanford University, the University of Sussex, and Texas A\&M University.}

This work was possible based on observations at NSF Cerro Tololo Inter-American Observatory, NSF NOIRLab (NOIRLab Prop. ID 2019A-0305; PI: Alex Drlica-Wagner), which is managed by the Association of Universities for Research in Astronomy (AURA) under a cooperative agreement with the U.S. National Science Foundation.

\CYTone{Some of the data presented herein were obtained at Keck Observatory, which is a private 501(c)3 non-profit organization operated as a scientific partnership among the California Institute of Technology, the University of California, and the National Aeronautics and Space Administration. The Observatory was made possible by the generous financial support of the W. M. Keck Foundation.}

\CYTone{The authors wish to recognize and acknowledge the very significant cultural role and reverence that the summit of Maunakea has always had within the Native Hawaiian community. We are most fortunate to have the opportunity to conduct observations from this mountain.}

This work has made use of data from the European Space Agency (ESA) mission
{\it Gaia} (\url{https://www.cosmos.esa.int/gaia}), processed by the {\it Gaia}
Data Processing and Analysis Consortium (DPAC,
\url{https://www.cosmos.esa.int/web/gaia/dpac/consortium}). Funding for the DPAC
has been provided by national institutions, in particular the institutions
participating in the {\it Gaia} Multilateral Agreement.

This work has made use of the Local Volume Database (\url{https://github.com/apace7/local_volume_database}) \citep{Pace2024arXiv241107424P}.

\nocite{Grillmair:2018}%\footnote{\CYTthree{A deep literature search revealed that Leo~VI also appeared among ${\sim}\,300$ possible candidates reported in an unpublished search of the PS1 data by .}}

\CYTone{This manuscript has been authored by Fermi Research Alliance, LLC, under contract No. DE-AC02-07CH11359 with the US Department of Energy, Office of Science, Office of High Energy Physics. The United States Government retains and the publisher, by accepting the article for publication, acknowledges that the United States Government retains a non-exclusive, paid-up, irrevocable, worldwide license to publish or reproduce the published form of this manuscript, or allow others to do so, for United States Government purposes}

%at Cerro Tololo InterAmerica Observatory at NSF’s NOIRLab, which is managed by the Association of Universities for Research in Astronomy (AURA) under a cooperative agreement with the National Science Foundation.

%\ADW{Remember to add standard DELVE acknowledgments.} CY: Added

%% in the research for the paper. Each keyword is check against the master 
%% list during copy editing. Individual instruments can be provided in 

\vspace{5mm}
\facilities{Blanco, Keck:II, Gaia}


\software{
\texttt{astropy}\citep{2013A&A...558A..33A,2018AJ....156..123A},
\texttt{corner} \citep{ForemanMackey:2016}, 
\texttt{emcee} \citep{Foreman_Mackey:2013},
\texttt{gala} \citep{Price-Whelan:2017},
\texttt{healpix} \citep{Gorski:2005},
\texttt{jupyter} \citep{Kluyver:16}, 
\texttt{matplotlib} \citep{Hunter:07}, 
\texttt{numpy} \citep{Oliphant:15}, 
\texttt{pandas}  \citep{McKinney:2010}, 
\texttt{scipy} \citep{Virtanen:2020},
\texttt{simple} \citep{Bechtol:2015}, 
\texttt{skyproj} (\url{https://github.com/LSSTDESC/skyproj}),
\texttt{ugali} \citep{Bechtol:2015,Drlica-Wagner:2021}, 
 }



%\appendix
%\section{Appendix information}
%Write Something


\bibliography{main}{}
\bibliographystyle{aasjournal}

%% the collaboration and author truncation commands are used. It has to

\appendix
\restartappendixnumbering
\section{\CYTthree{Posterior Distribution of the Leo VI Parameters } }
\CYTthree{In Figure \ref{Figure:AppendixA}, we present the posterior distribution of a simultaneous fit to Leo VI's morphological and isochrone properties derived from follow-up DECam observations of the system using \texttt{ugali}, as described in Section \ref{sec:fits}. The posterior was sampled with \texttt{emcee} using 40 walkers over 3000 steps, following an initial burn-in of 1000 steps.}

\begin{figure*}[ht]
\centering
\includegraphics[width=0.92\linewidth]{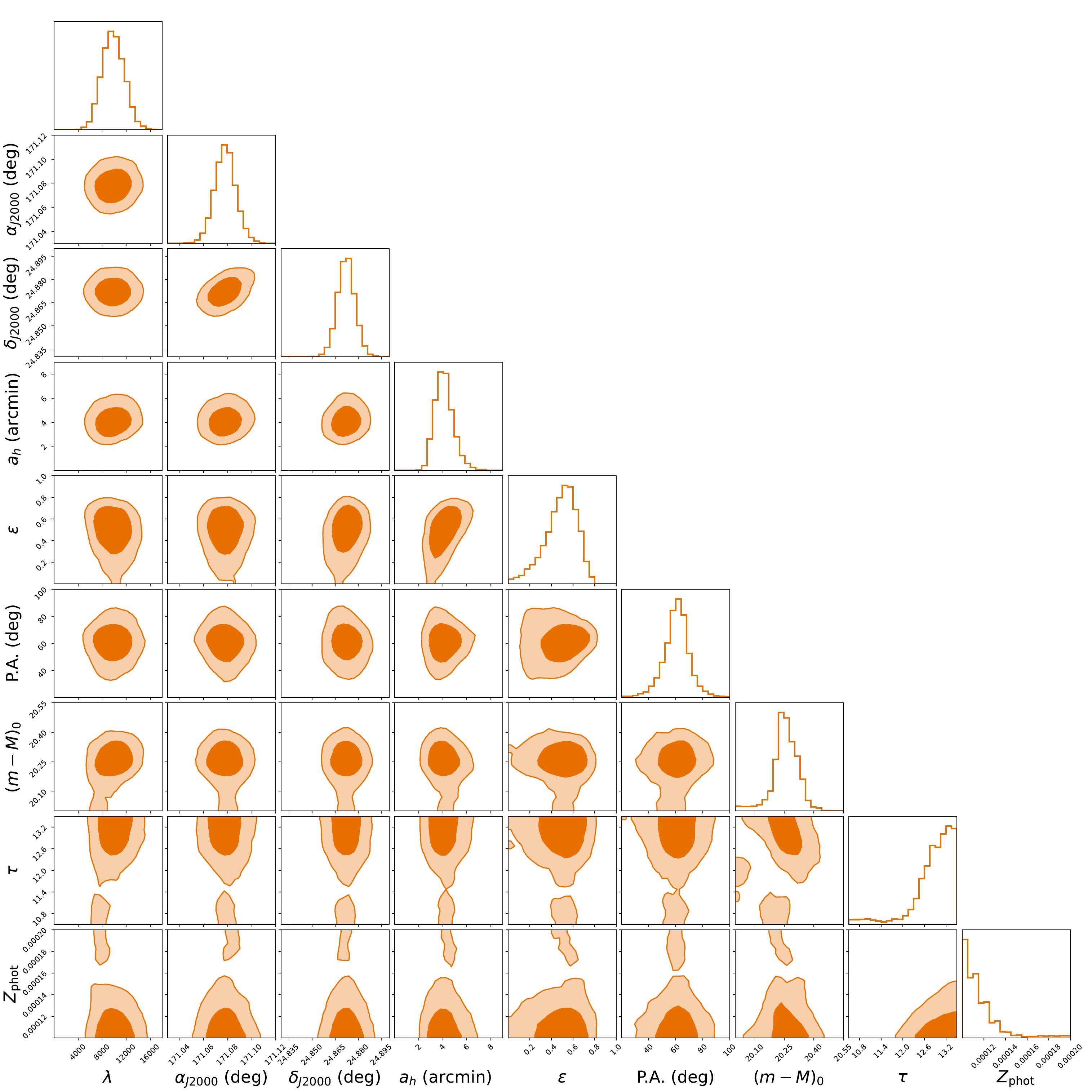}
%\includegraphics[width=\linewidth]{plots/Fig1.pdf}

%\vspace{-3cm}
\caption{\CYTthree{The posterior distribution of Leo VI's morphological and isochrone properties modeled using \texttt{ugali}. We simultaneously fit the system's stellar richness ($\lambda$), centroid coordinates ($\alpha_{J2000}, \delta_{J2000}$), angular semi-major axis length ($a_h$), ellipticity ($\epsilon$),  position angle (P.A.) of the major axis, distance modulus ($(m-M)_0$),  age ($\tau$), and metallicity ($Z_{\rm phot}$). } \label{Figure:AppendixA}}
\end{figure*}

\CYTthree{We present the  posterior probability distributions for Leo VI's heliocentric radial velocity, $v_{hel}$, and velocity dispersion, $\sigma_{v}$ (Figure \ref{Figure:Keckcorner_vel}) and the distributions for its  spectroscopic metallicity, $\mbox{[Fe/H]}_{\rm spec}$, and metallicity dispersion, $\sigma_{\rm [Fe/H]}$ (Figure \ref{Figure:Keckcorner_feh}),  obtained using data from the DEIMOS observation of the system (see Section \ref{sec:Keck}). }

\begin{figure*}[ht]
\centering
\includegraphics[width=0.46\linewidth]{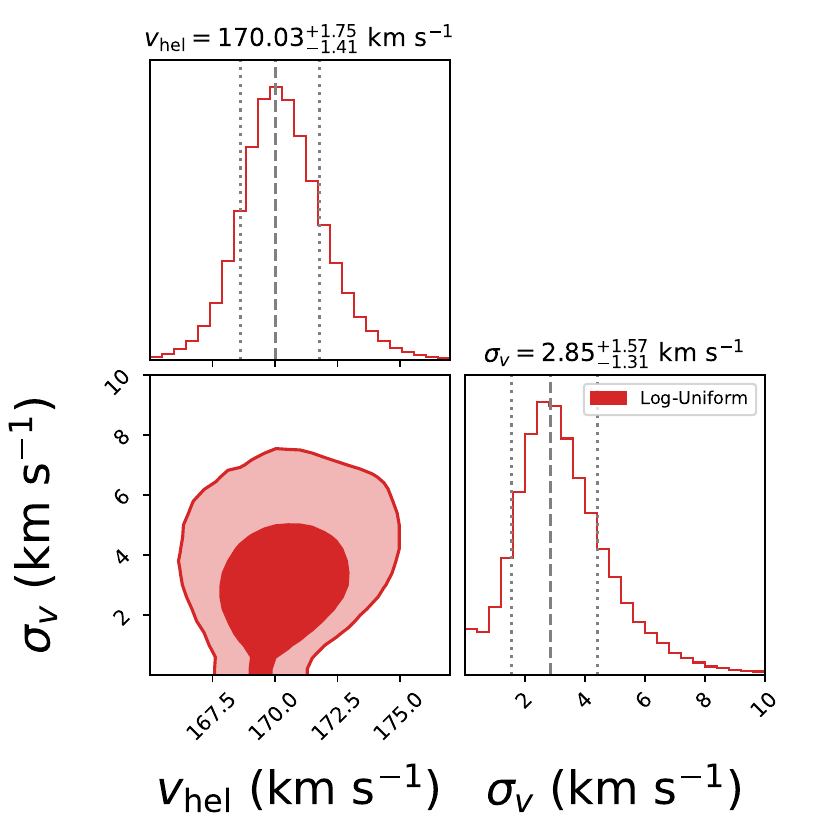}
\includegraphics[width=0.46\linewidth]{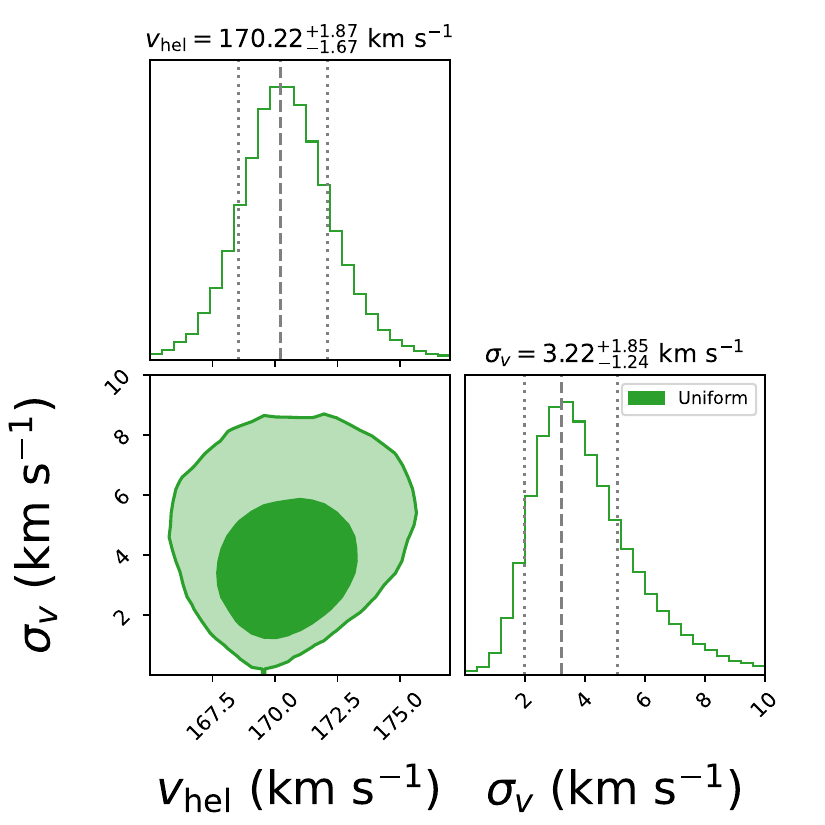}
%\vspace{-3cm}
\caption
{\CYTthree{The posterior probability distributions for the systemic heliocentric radial velocity, $v_{hel}$, and velocity dispersion, $\sigma_{v}$, of Leo VI derived using MCMC sampling. The left plot shows values obtained using a log-uniform prior while the right plot shown alternative values obtained using a uniform prior. }}
\label{Figure:Keckcorner_vel}
\end{figure*}

\begin{figure*}[ht]
\centering
\includegraphics[width=0.46\linewidth]{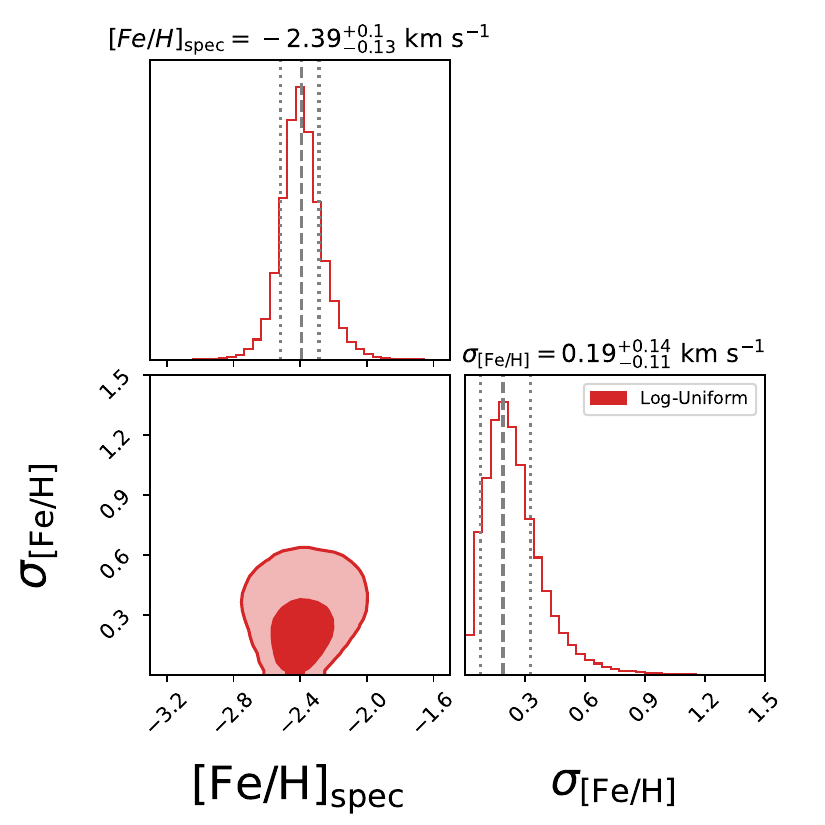}
\includegraphics[width=0.46\linewidth]{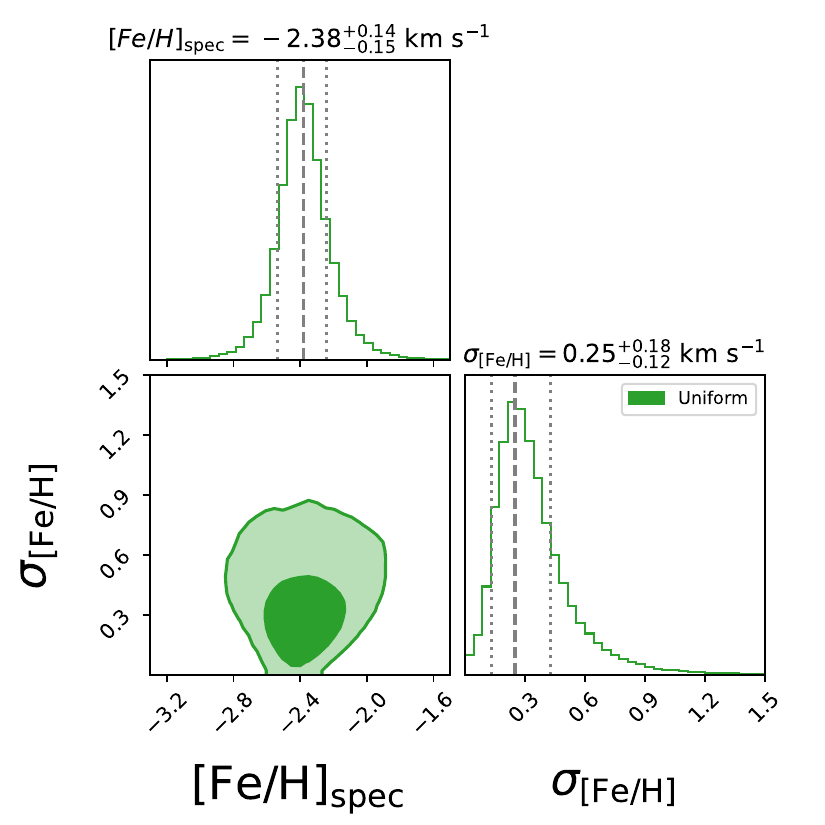}
%\vspace{-3cm}
\caption
{\CYTthree{{The posterior probability distributions for the  spectroscopic metallicity, $\mbox{[Fe/H]}_{\rm spec}$, and metallicity dispersion, $\sigma_{\rm [Fe/H]}$, of the system.  The left plot shows values obtained using a log-uniform prior while the right plot shown alternative values obtained using a uniform prior.}}}
\label{Figure:Keckcorner_feh}
\end{figure*}

\CYTthree{Figure \ref{Figure:orbit_chains} shows the distribution of Leo VI's systematic proper motion and orbital parameters estimated from \textit{Gaia} proper motion measurements of Leo VI's member stars (Section \ref{sec:orbit})}.

\begin{figure*}[ht]
\centering
\includegraphics[width=0.95\linewidth]{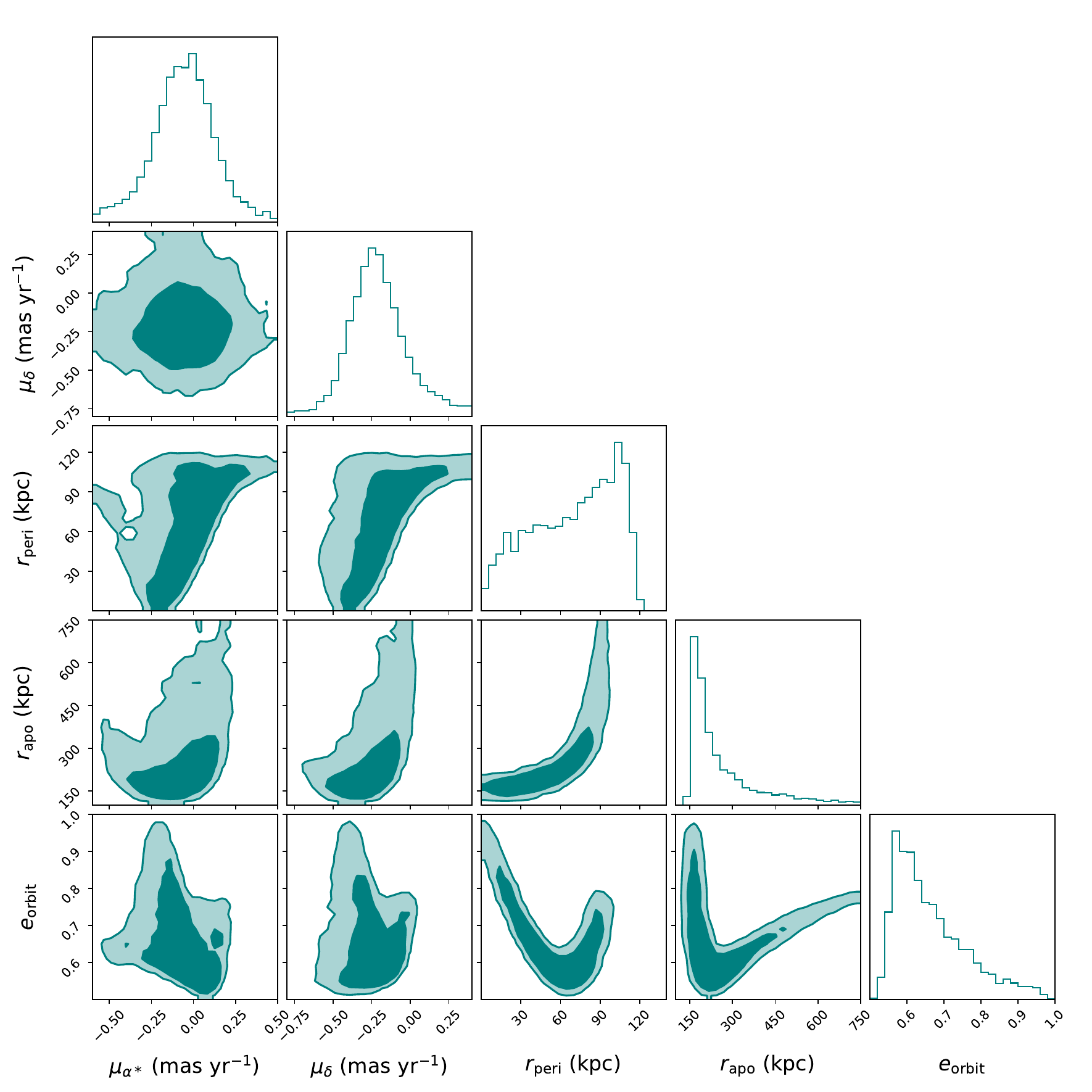}
%\includegraphics[width=\linewidth]{plots/Fig1.pdf}

%\vspace{-3cm}
\caption{\CYTthree{The distribution of Leo VI's systematic proper motion and orbital parameters. The systematic proper motion ($\mu_{\alpha * }$, $\mu_{\delta} $) were obtained from the proper motion measurements of Leo VI's member stars using a Gaussian mixture model sampled with \texttt{emcee}. While the orbital properties such as apocenter ($r_{\rm apo}$), pericenter ($r_{\rm peri}$) and eccentricity ($e_{\rm orbit}$) were obtained using \texttt{gala} with the distribution of the present-day 6-D position and velocities obtained by sampling the posterior distribution of the system's observed parameters. \label{Figure:orbit_chains}}}
\end{figure*}

\end{document}